\documentclass[conference]{IEEEtran}
\usepackage{graphicx}
\usepackage{amssymb}
\usepackage{pifont}
\usepackage[all=normal, title=normal, paragraphs=tight, wordspacing=normal]{savetrees}
\usepackage{comment}
\let\labelindent\relax
\usepackage{framed}
\usepackage{balance}
\usepackage{amsmath}
\usepackage{booktabs} 
\usepackage{xcolor}
\usepackage{xspace}
\usepackage{microtype}
\usepackage{times}
\usepackage[noend]{algpseudocode}
\usepackage{amssymb}
\usepackage{subfig}
\usepackage{algpseudocode}
\usepackage{listings}
\usepackage{algorithm}
\usepackage{url}
\usepackage{multirow} 
\usepackage{listings}
\usepackage{color}
\usepackage{bm}
\usepackage{microtype} 
\usepackage{enumitem}
\usepackage{framed}
\usepackage{float}

\usepackage[turnon]{notes}
\usepackage{multirow}
\usepackage{xspace}
\usepackage{listings}
\usepackage[noabbrev]{cleveref}

\usepackage[utf8]{inputenc}
\usepackage[T1]{fontenc}
\usepackage{microtype}

\renewcommand{\baselinestretch}{0.99}

\newcommand{\rom}[1]{\uppercase\expandafter{\romannumeral #1\relax}}

\newcommand{\eg}{\hbox{\emph{e.g.}}\xspace}
\newcommand{\ie}{\hbox{\emph{i.e.}}\xspace}

\newcommand{\wrt}{\hbox{\emph{w.r.t.}}\xspace}

\definecolor{gray50}{gray}{.5}
\definecolor{gray40}{gray}{.6}
\definecolor{gray30}{gray}{.7}
\definecolor{gray20}{gray}{.8}
\definecolor{gray10}{gray}{.9}
\definecolor{gray05}{gray}{.95}

\newlength\Linewidth
\def\findlength{\setlength\Linewidth\linewidth
\addtolength\Linewidth{-4\fboxrule}
\addtolength\Linewidth{-3\fboxsep}
}
\newenvironment{examplebox}{\par\begingroup
   \setlength{\fboxsep}{5pt}\findlength
   \setbox0=\vbox\bgroup\noindent
   \hsize=0.95\linewidth
   \begin{minipage}{0.95\linewidth}\normalsize}
    {\end{minipage}\egroup
    \textcolor{gray20}{\fboxsep1.5pt\fbox
     {\fboxsep5pt\colorbox{gray05}{\normalcolor\box0}}}
    \endgroup\par\noindent
    \normalcolor\ignorespacesafterend}

\newcounter{RQCounter}
\newcounter{RQACounter}

\newcommand{\RQA}[2]{%
\refstepcounter{RQACounter} \label{#1}
\vspace{0.1in} 
\subsection*{\hspace{0.05cm}\textbf{RQ\arabic{RQACounter}.~#2} \vspace{0.05in}}

}

\newcommand{\RS}[2]{%
\begin{framed}%
\filbreak
\textbf{Result {\ref{#1}}:~}{\emph {#2}}%
\end{framed}
}

\definecolor{javared}{rgb}{0.6,0,0} 
\definecolor{javagreen}{rgb}{0.25,0.5,0.35} 
\definecolor{javapurple}{rgb}{0.5,0,0.35} 
\definecolor{javadocblue}{rgb}{0.25,0.35,0.75} 

\lstdefinestyle{customc}{
  belowcaptionskip=\baselineskip,
  breaklines=true,
  xleftmargin=\parindent,
  language=java,
  showstringspaces=false,
  basicstyle=\footnotesize\ttfamily,
  keywordstyle=\bfseries\color{javapurple},
  commentstyle=\itshape\blue,
 numbers=left,
 numbersep=.1cm, 
 linewidth=\columnwidth,
 xleftmargin=2.2em,
 framexleftmargin=1.1em,
 frame=single,
 float=H, 
 aboveskip=\baselineskip
}

\lstdefinestyle{CStyle}{
    backgroundcolor=\color{backgroundColour},   
    commentstyle=\color{mGreen},
    keywordstyle=\color{magenta},
    numberstyle=\tiny\color{mGray},
    stringstyle=\color{mPurple},
    basicstyle=\footnotesize,
    breakatwhitespace=false,         
    breaklines=true,                 
    captionpos=b,                    
    keepspaces=true,                 
    numbers=left,                    
    numbersep=5pt,                  
    showspaces=false,                
    showstringspaces=false,
    showtabs=false,                  
    tabsize=2,
    language=C
}

\definecolor{codegreen}{rgb}{0,0.6,0}
\definecolor{codegray}{rgb}{0.5,0.5,0.5}
\definecolor{codepurple}{rgb}{0.58,0,0.82}
\definecolor{backcolour}{rgb}{0.95,0.95,0.92}

\lstdefinestyle{braystyle}{
  commentstyle=\Red,
  keywordstyle=\blue,
  numberstyle=\tiny\color{codegray},
  stringstyle=\color{codepurple},
  basicstyle=\scriptsize,
  breakatwhitespace=false,         
  breaklines=true,                 
  captionpos=b,                    
  keepspaces=true,                 
  numbers=left,                    
  numbersep=4pt,                  
  showspaces=false,                
  showstringspaces=false,
  showtabs=false,                  
  tabsize=2,
  belowskip=-3
  \baselineskip,
  frame=None,
  language=C,
  float=H
}

\DeclareMathOperator*{\argmin}{arg\,min}

\definecolor{mGreen}{rgb}{0,0.6,0}
\definecolor{mGray}{rgb}{0.5,0.5,0.5}
\definecolor{mPurple}{rgb}{0.58,0,0.82}
\definecolor{backgroundColour}{rgb}{0.95,0.95,0.92}

\newcommand{\afl}{AFL\xspace}
\newcommand{\aflfast}{AFLFast\xspace}
\newcommand{\vuzzer}{VUzzer\xspace}
\newcommand{\kleefl}{KleeFL\xspace}
\newcommand{\lafintel}{AFL-laf-intel\xspace}

\newcommand{\sys}{\textsc{Neuzz}\xspace}

\newcommand{\rqa}{Can \sys find more bugs than existing fuzzers?}
\newcommand{\rqb}{Can \sys achieve higher edge coverage than existing fuzzers?}
\newcommand{\rqc}{Can \sys perform better than existing RNN-based fuzzers?}
\newcommand{\rqd}{How do different model choices affect \sys's performance?}

\newcommand\Red[1]{\textcolor[rgb]{1.00,0.00,0.00}{\textbf{#1}}}

\newcommand\blue[1]{\textcolor[rgb]{0.00,0.00,1.00}{{#1}}}

\begin{document}
\title{NEUZZ: Efficient Fuzzing with Neural \\ Program Smoothing}
\author{\IEEEauthorblockN{Dongdong She,
Kexin Pei, Dave Epstein, Junfeng Yang, Baishakhi Ray, and 
Suman Jana}
\IEEEauthorblockA{Columbia University\\}
}

\maketitle
\thispagestyle{plain}
\pagestyle{plain}

\begin{abstract}
Fuzzing has become the de facto standard technique for finding software vulnerabilities. However, even  state-of-the-art fuzzers are not very efficient at finding hard-to-trigger software bugs. Most popular fuzzers use evolutionary guidance to generate inputs that can trigger different bugs. Such evolutionary algorithms, while fast and simple to implement, often get stuck in fruitless sequences of random mutations. Gradient-guided optimization presents a promising alternative to evolutionary guidance. Gradient-guided techniques have been shown to significantly outperform evolutionary algorithms at solving high-dimensional structured optimization problems in domains like machine learning by efficiently utilizing gradients or higher-order derivatives of the underlying function. 

However, gradient-guided approaches are not directly applicable to fuzzing as real-world program behaviors contain many discontinuities, plateaus, and ridges where the gradient-based methods often get stuck. We observe that this problem can be addressed by creating a smooth surrogate function approximating the target program’s discrete branching behavior. In this paper, we propose a novel program smoothing technique using surrogate neural network models that can incrementally learn smooth approximations of a complex, real-world program's branching behaviors. We further demonstrate that such neural network models can be used together with gradient-guided input generation schemes to significantly increase the efficiency of the fuzzing process. 

Our extensive evaluations demonstrate that NEUZZ significantly outperforms 10 state-of-the-art graybox fuzzers on 10 popular real-world programs both at finding new bugs and achieving higher edge coverage. NEUZZ found 31 previously unknown bugs (including two CVEs) that other fuzzers failed to find in 10 real-world programs and achieved 3X more edge coverage than all of the tested graybox fuzzers over 24 hour runs. Furthermore, NEUZZ also outperformed existing fuzzers on both LAVA-M and DARPA CGC bug datasets.
\end{abstract}


\section{Introduction}
\label{sec:intro}

Fuzzing has become the de facto standard technique for finding software vulnerabilities~\cite{afl, cgc}. The fuzzing process involves generating random test inputs and executing the target program with these inputs to trigger potential security vulnerabilities~\cite{miller1990empirical}. Due to its simplicity and low performance overhead, fuzzing has been very successful at finding different types of security vulnerabilities in many real-world programs~\cite{ossfuzz, clusterfuzz, duchene2014kameleonfuzz, slowfuzz, aflfast, libfuzzer}. Despite their tremendous promise, popular fuzzers, especially for large programs, often tend to get stuck trying redundant test inputs and struggle to find security vulnerabilities hidden deep within program logic~\cite{driller, sage, pengt}. 

Conceptually, fuzzing is an optimization problem whose goal is to find program inputs that maximize the number of vulnerabilities found within a given amount of testing time~\cite{miller2007fuzz}. However, as security vulnerabilities tend to be sparse and erratically distributed across a program, most fuzzers aim to test as much program code as they can by maximizing some form of code coverage (\eg, edge coverage) to increase their chances of finding security vulnerabilities. Most popular fuzzers use evolutionary algorithms to solve the underlying optimization problem\textemdash generating new inputs that maximize code coverage~\cite{afl, aflfast, libfuzzer, zzuf}. Evolutionary optimization starts from a set of seed inputs, applies random mutations to the seeds to generate new test inputs, executes the target program for these inputs, and only keeps the promising new inputs (\eg, those that achieve new code coverage) as part of a corpus for further mutation. However, as the input corpus gets larger, the evolutionary process becomes increasingly less efficient at reaching new code locations. 

One of the main limitations of evolutionary optimization algorithms is that they cannot leverage the structure (\ie, gradients or other higher-order derivatives) of the underlying optimization problem. Gradient-guided optimization (\eg, gradient descent) is a promising alternative approach that has been shown to significantly outperform evolutionary algorithms at solving high-dimensional structured optimization problems in diverse domains including aerodynamic computations and machine learning~\cite{zingg2008comparative, horst2013handbook, goodfellow2016deep}. 

However, gradient-guided optimization algorithms cannot be directly applied to fuzzing real-world programs as they often contain significant amounts of discontinuous behaviors (cases where the gradients cannot be computed accurately) due to widely different behaviors along different program branches~\cite{parnas1985software, chaudhuri2011smoothing, harman2010theoretical,chaudhuri2010smooth,angora}. We observe that this problem can be overcome by creating a smooth (\ie, differentiable) surrogate function approximating the target program's branching behavior with respect to program inputs. Unfortunately, existing program smoothing techniques~\cite{chaudhuri2011smoothing, chaudhuri2010smooth} incur prohibitive performance overheads as they depend heavily on symbolic analysis that does not scale to large programs due to several fundamental limitations like path explosion, incomplete environment modeling, and large overheads of symbolic memory modeling~\cite{king1976symbolic, sen2005cute, cadar2008klee,cadar2013symbolic,cadar2011symbolic,godefroid2005dart, khurshid2003generalized}. 

In this paper, we introduce a novel, efficient, and scalable program smoothing technique using feed-forward Neural Networks (NNs) that can incrementally learn smooth approximations of complex, real-world program branching behaviors, \ie, predicting the control flow edges of the target program exercised by a particular given input. We further propose a gradient-guided search strategy that computes and leverages the gradient of the smooth approximation (\ie, an NN model) to identify target mutation locations that can maximize the number of detected bugs in the target program. We demonstrate how the NN model can be refined by incrementally retraining the model on mispredicted program behaviors. We find that feed-forward NNs are a natural fit for our task because of (i) their demonstrated ability to approximate complex non-linear functions, as implied by the universal approximation theorem~\cite{funahashi1989approximate}, and (ii) their support for efficient and accurate computation of gradients/higher-order derivatives~\cite{goodfellow2016deep}. 

We design and implement our technique as part of \sys, a new learning-enabled fuzzer. We compare \sys with $10$ state-of-the art fuzzers on $10$ real-world programs covering $6$ different file formats, (\eg, ELF, PDF, XML, ZIP, TTF, and JPEG) with an average of $47,546$ lines of code, the LAVA-M bug dataset~\cite{lava}, and the CGC dataset~\cite{cgc_repo}. Our results show that \sys consistently outperforms all the other fuzzers by a wide margin both in terms of detected bugs and achieved edge coverage. \sys found $31$ previously unknown bugs (including CVE-2018-19931 and CVE-2018-19932) in the tested programs that other fuzzers failed to find. Our tests on the DARPA CGC dataset also confirmed that \sys can outperform state-of-the-art fuzzers like Driller~\cite{driller} at finding different bugs.

Our primary contributions in this paper are as follows:

\begin{itemize}
\item 
We are the first to identify the significance of program smoothing for adopting efficient gradient-guided techniques for fuzzing.  

\item 
We introduce the first efficient and scalable program smoothing technique using surrogate neural networks to effectively model the target program's branching behaviors. We further propose an incremental learning technique to iteratively refine the surrogate model as more training data becomes available.

\item
We demonstrate that the gradients of the surrogate neural network model can be used to efficiently generate program inputs that maximize the number of bugs found in the target program. 

\item 
We design, implement, and evaluate our techniques as part of \sys and demonstrate that it significantly outperforms 10 state-of-the-art fuzzers on a wide range of real-world programs as well as curated bug datasets.
\end{itemize}

The rest of the paper is organized as follows. Section~\ref{sec:overview} summarizes the necessary background information on optimization and gradient-guided techniques. Section~\ref{subsec:fuzz_learn} provides an overview of our technique along with a motivating example. Section~\ref{sec:methodology} and Section~\ref{sec:impl} describe our methodology and implementation in detail. We present our experimental results in Section~\ref{sec:eval} and describe some sample bugs found by \sys in Section~\ref{sec:case_study}. Section~\ref{sec:related} summarizes the related work and Section~\ref{sec:conclusion} concludes the paper.

\section{Optimization Basics}
\label{sec:overview}
In this section, we first describe the basics of optimization and the benefits of gradient-guided optimization over evolutionary guidance for smooth functions. 
Finally, we demonstrate how fuzzing can be cast as an optimization problem. 

An optimization problem usually consists of three different components: a vector of parameters $x$, an objective function $F(x)$ to be minimized or maximized, and a set of constraint functions $C_i(x)$ each involving either inequality or equality that must be satisfied. The goal of the optimization process is to find a concrete value of the parameter vector $x$ that maximizes/minimizes $F(x)$ while satisfying all constraint functions $C_i(x)$ as shown below. 

\begin{equation}
\centering
\underset{x \in R^n}{max/min}\textrm{    }F(x) \textrm{    subject to    }
\begin{cases} 
      C_i(x)\geq 0, i \in N \\
      C_i(x) = 0, i \in Q \\
\end{cases}
\end{equation}

Here $R$, $N$, and $Q$ denote the sets of real numbers, the indices for inequality constraints, and the indices for equality constraints, respectively.

\vspace{0.2cm}
\noindent
{\bf Function smoothness \& optimization.}
Optimization algorithms usually operate in a loop beginning with an initial guess of the parameter vector $x$ and gradually iterating to find better solutions. The key component of any optimization algorithm is the strategy it uses to move from one value of $x$ to the next. Most strategies leverage the values of the objective function $F$, the constraint functions $C_i$, and, if available, the gradient/higher-order derivatives. 

The ability and efficiency of different optimization algorithms to converge to the optimal solution heavily depend on the nature of the objective and constraint functions  $F$ and $C_i$. In general, smoother functions (\ie, those with well-defined and computable derivatives) can be more efficiently optimized than functions with many discontinuities (\eg, ridges or plateaus). Intuitively, the smoother the objective/constraint functions are, the easier it is for the optimization algorithms to accurately compute gradients or higher-order derivatives and use them to systematically search the entire parameter space. 

\begin{figure}
\centering
\subfloat[{\bf gradient descent}]{\includegraphics[width=0.23\textwidth]{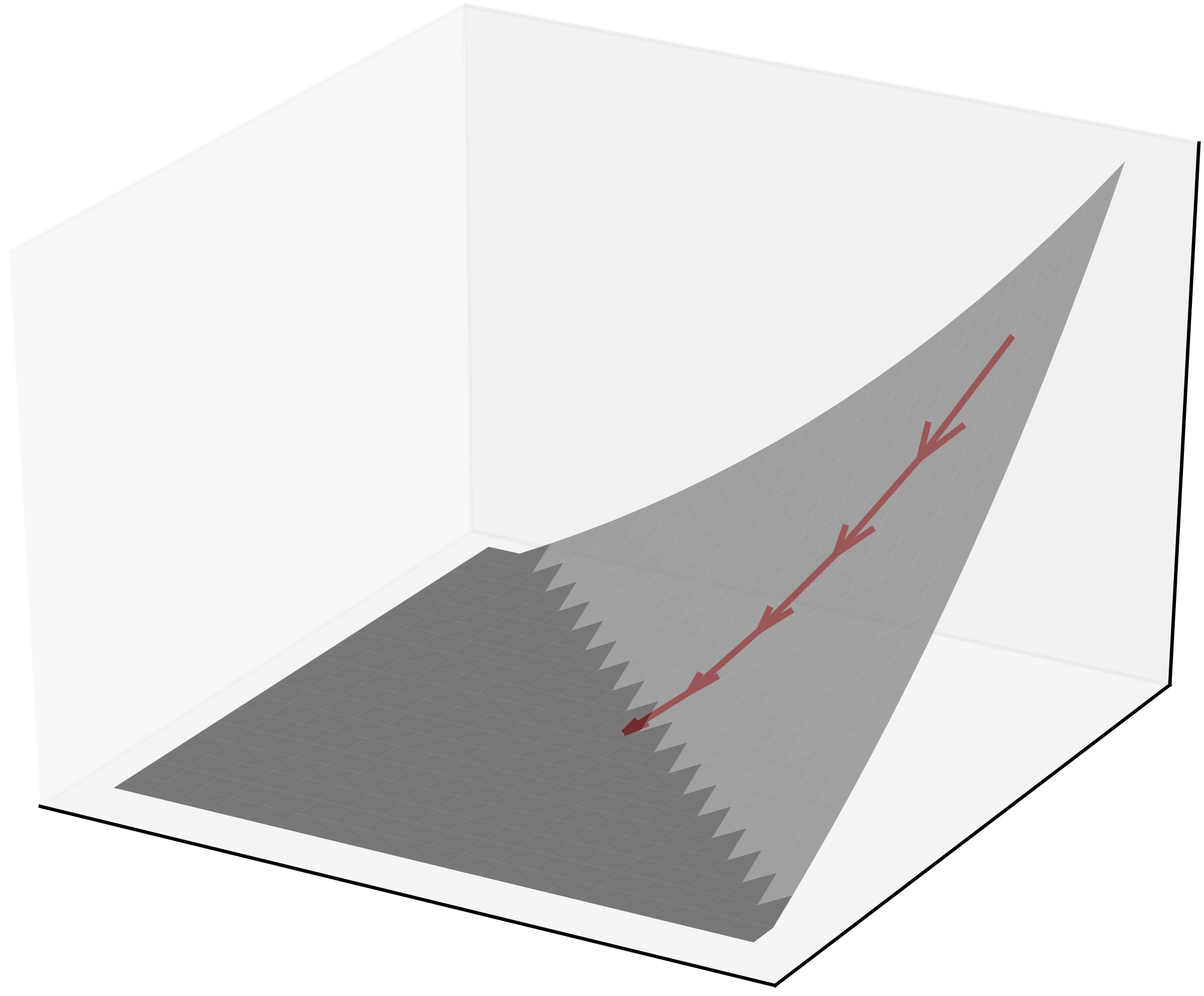}\label{gd}}
\hspace{.2cm}
	\subfloat[{\bf evolutionary algorithm}]{\includegraphics[width=0.23\textwidth]{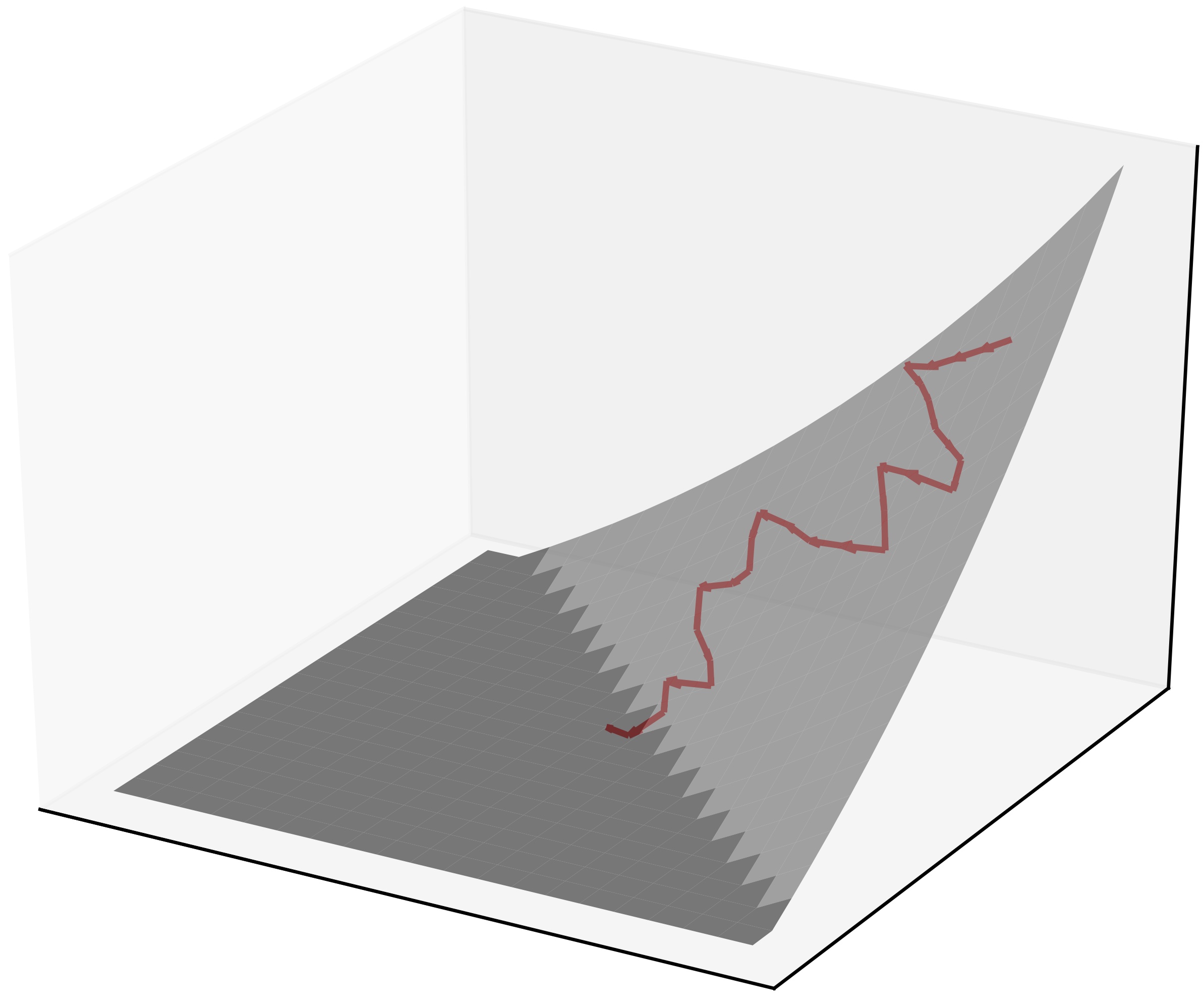}\label{ga}}
\caption{{\bf\small Gradient-guided optimization algorithms like gradient descent can be significantly more efficient than evolutionary algorithms that do not use any gradient information}}
\label{gd_ga}
\end{figure}

For the rest of this paper, we specifically focus on unconstrained optimization problems that do not have any constraint functions, \ie,  $C=\phi$, as they closely mimic fuzzing, our target domain. For unconstrained smooth optimization problems, gradient-guided approaches can significantly outperform evolutionary strategies at solving high-dimensional structured optimization problems ~\cite{zingg2008comparative, horst2013handbook, goodfellow2016deep}. This is because gradient-guided techniques effectively leverage gradients/higher-order derivatives to efficiently converge to the optimal solution as shown in Figure~\ref{gd_ga}. 

\vspace{0.2cm}
\noindent
{\bf Convexity \& gradient-guided optimization.} For a common class of functions called convex functions, gradient-guided techniques are highly efficient and can always converge to the globally optimal solution~\cite{wright1999numerical}.  Intuitively, a function is convex if a straight line connecting any two points on the graph of the function lies entirely above or on the graph. More formally, a function f is called  convex if the following property is satisfied by all pairs of points $x$ and $y$ in its domain: $f (tx + (1 - t)y) \leq t f(x) + (1 - t) f(y), \forall t \in [0, 1]$.  

However, in non-convex functions, gradient-guided approach may get stuck at locally optimal solutions where the objective function is greater (assuming that the goal is to maximize) than all nearby feasible points but there are other larger values present elsewhere in the entire range of feasible parameter values.  However, even for such cases, simple heuristics like restarting the gradient-guided methods from new randomly chosen starting points have been shown to be highly effective in practice~\cite{goodfellow2016deep, wright1999numerical}. 
 
\vspace{0.2cm}
\noindent
{\bf Fuzzing as unconstrained optimization.}
Fuzzing can be represented as an unconstrained optimization problem where the objective is to maximize the number of bugs/vulnerabilities found in the test program for a fixed number of test inputs. Therefore, the objective function can be thought of as $F_p(x)$, which returns $1$ if input $x$ triggers a bug/vulnerability when the target program $p$ is executed with input $x$.  However, such a function is too ill-behaved (\ie, mostly containing flat plateaus and a few very sharp transitions) to be optimized efficiently. 

Therefore, most graybox fuzzers instead try to maximize the amount of tested code (\eg, maximize edge coverage) as a stand-in proxy metric~\cite{afl, aflfast, vuzzer, steelix, angora}.  Such an objective function can be represented as $F'_p(x)$ where $F'$ returns the number of new control flow edges covered by the input x for program $P$. Note that $F'$ is relatively easier to optimize than the original function $F$ as the number of all possible program inputs exercising new control flow edges tend to be significantly higher than the inputs that trigger bugs/security vulnerabilities. 

Most existing graybox fuzzers use evolutionary techniques~\cite{afl, aflfast, vuzzer, steelix, angora} along with other domain-specific heuristics as their main optimization strategy. The key reason behind picking such algorithms over gradient-guided optimization is that most real-world programs contain many discontinuities due to significantly different behaviors along different program paths~\cite{chaudhuri2012continuity}. Such discontinuities may cause the gradient-guided optimization to get stuck at non-optimal solutions. In this paper, we propose a new technique using a neural network for smoothing the target programs to make them suitable for gradient-guided optimization and demonstrate how fuzzers might exploit such strategies to significantly boost their effectiveness.

\begin{figure}[!t]
\centering
\includegraphics[width=\columnwidth]{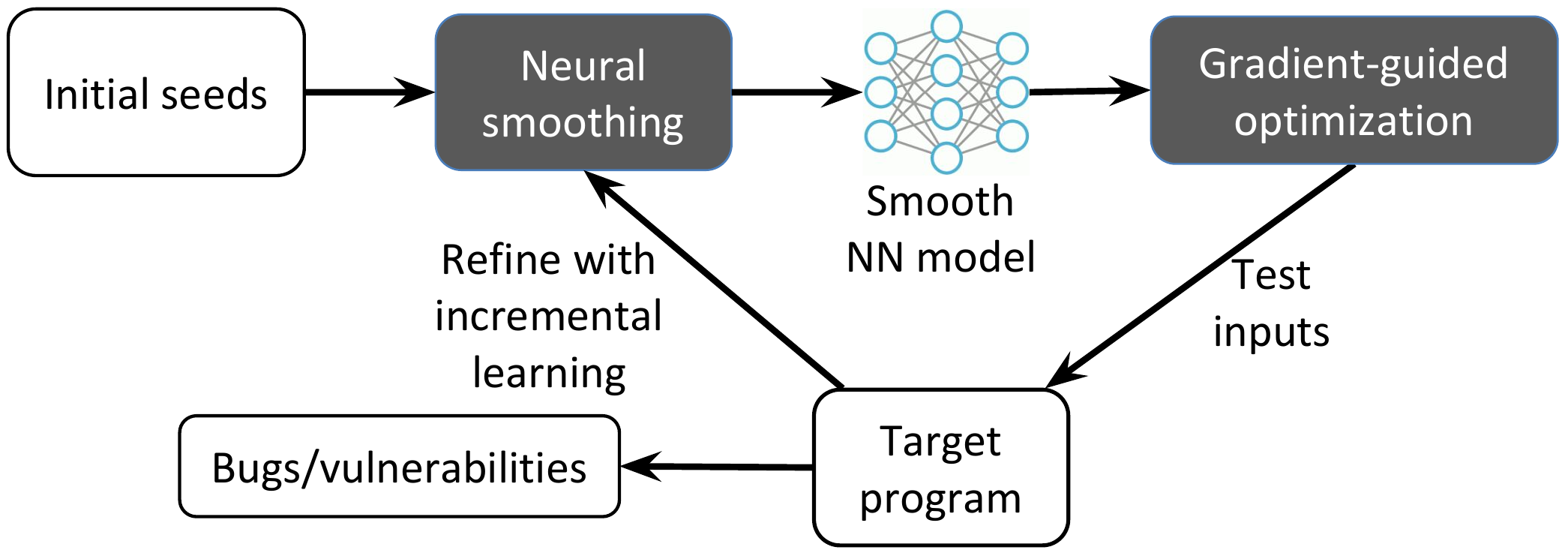}
\caption{\textbf{\small An overview of our approach}}
\label{fig:workflow_overview}
\end{figure}

\begin{figure*}[!t]
\centering
\begin{tabular}{cc}
\begin{minipage}[t]{0.7\linewidth}
\centering
\subfloat[Original]{\includegraphics[width=0.33\linewidth]{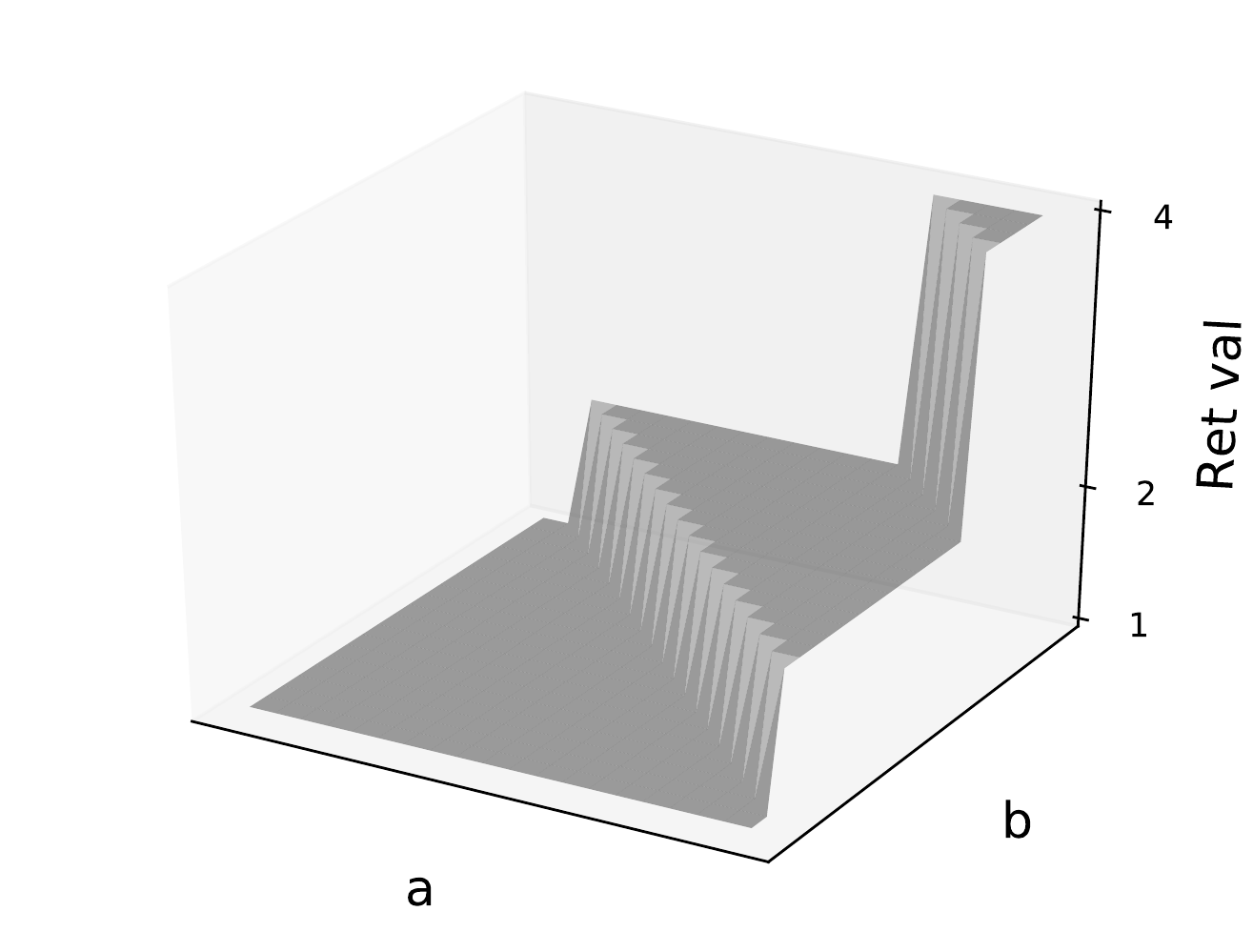}\label{subfig:unsmooth}}
\subfloat[NN smoothing]{\includegraphics[width=0.33\linewidth]{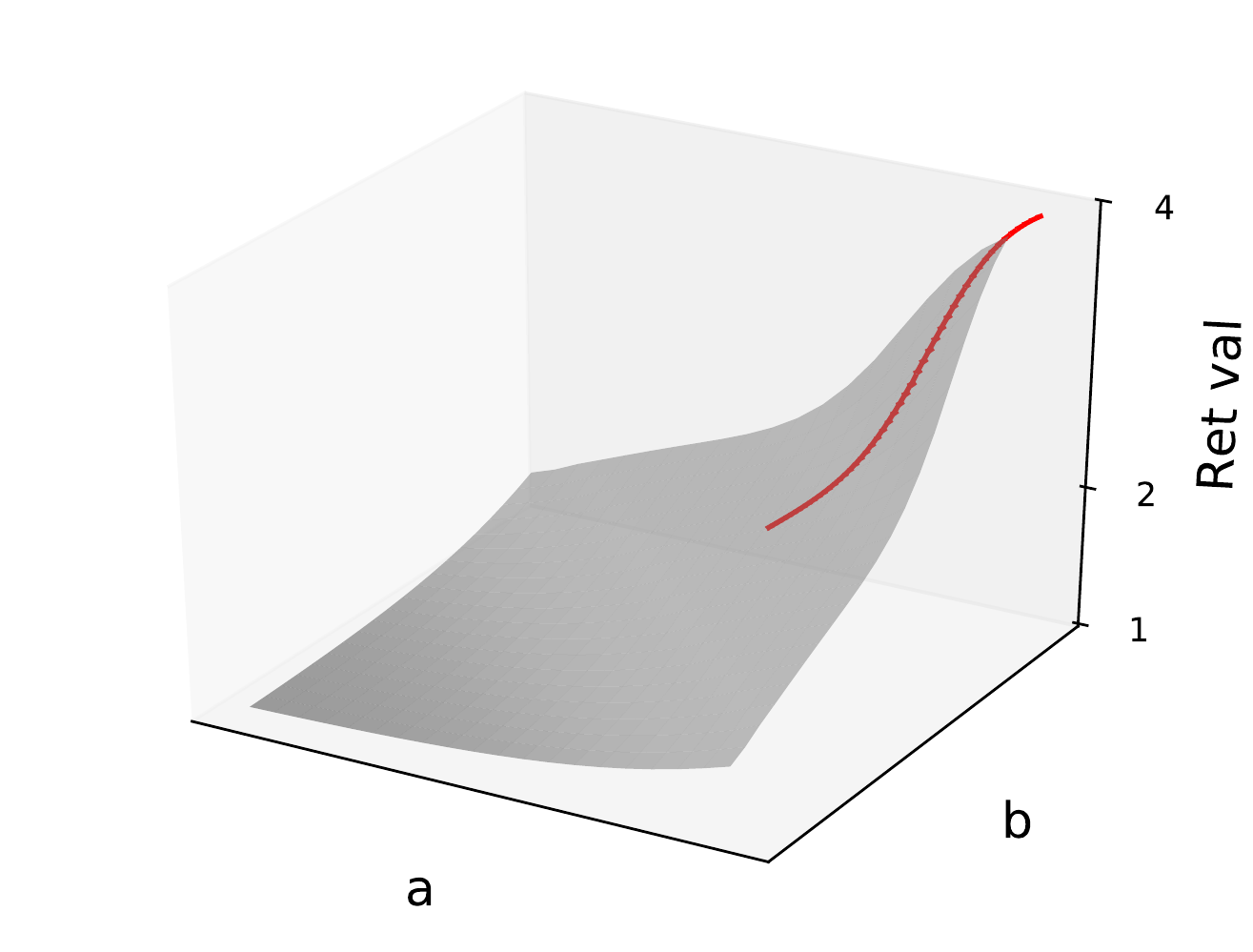}\label{subfig:smoother}}
\subfloat[NN smoothing + refining]{\includegraphics[width=0.33\linewidth]{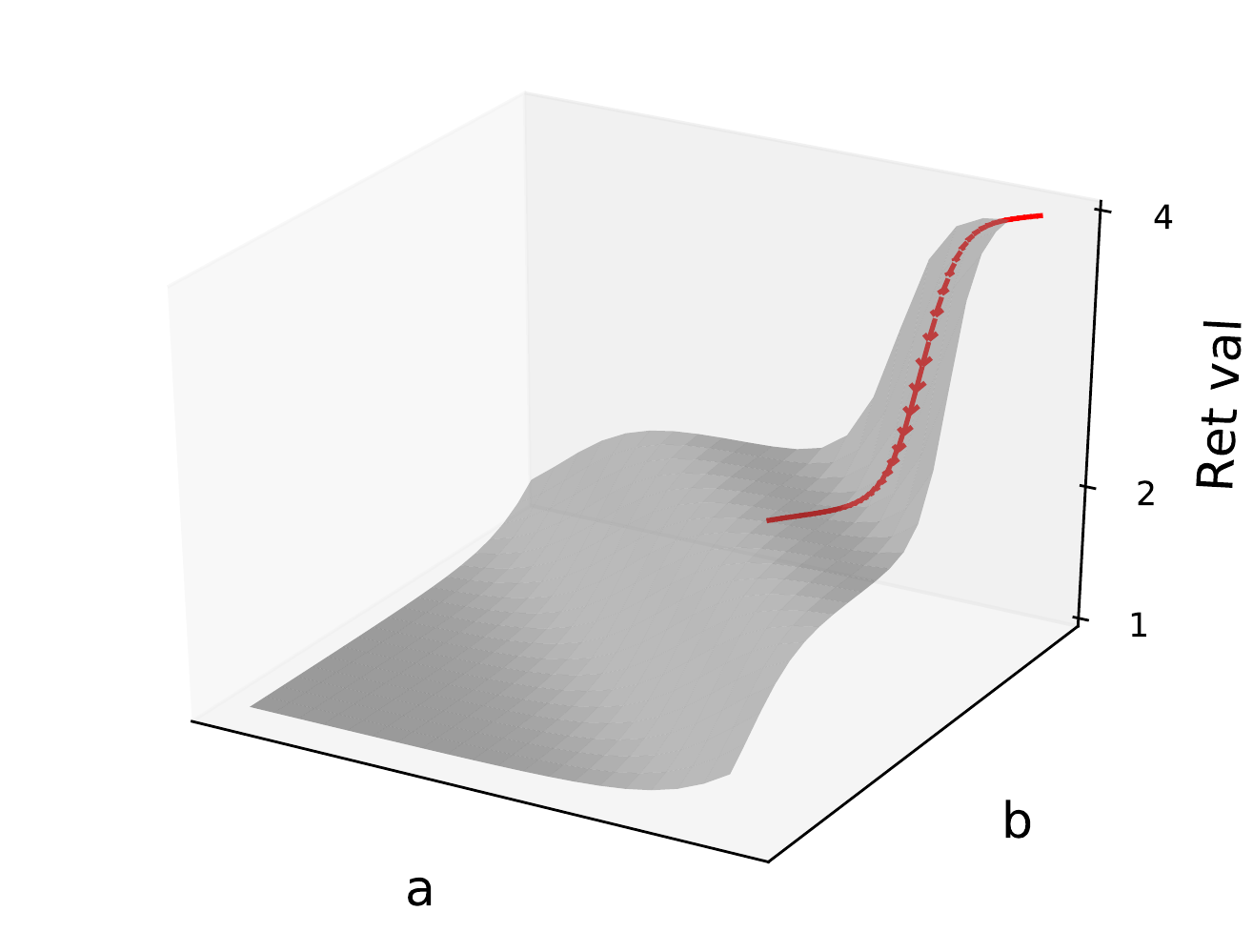}\label{subfig:smoothest}}
\end{minipage}
&
\begin{minipage}[b]{.3\linewidth}
\lstset{basicstyle=\footnotesize\ttfamily,breaklines=true}
\begin{lstlisting} %[xleftmargin=1cm,frame=single]
 
z = pow(3, a+b);
if(z < 1){
  return 1;
} 
else if(z < 2){
  //vulnerability
  return 2;
}  
else if(z < 4){ 
  return 4;
}  
\end{lstlisting} 
\end{minipage}
\\
\end{tabular}
\vspace{0.4cm}
\caption{\textbf{\small Simple code snippet demonstrating the benefits of neural smoothing for fuzzing}}
\label{fig:workflow}
\end{figure*}

\section{Overview of Our Approach}
\label{subsec:fuzz_learn}
Figure~\ref{fig:workflow_overview} presents a high level overview of our approach. We describe the key components in detail below.

\vspace{0.2cm}
\noindent\textbf{Neural program smoothing.} 
Approximating a program's discontinuous branching behavior smoothly is essential for accurately computing gradients or higher-order derivatives that are necessary for gradient-guided optimization. Without such smoothing, the gradient-guided optimization process may get stuck at different discontinuities/plateaus. The goal of the smoothing process is to create a smooth function that can mimic a program's branching behavior without introducing large errors (\ie, it deviates minimally from the original program behavior). We use a feed-forward neural network (NN) for this purpose. As implied by the universal approximation theorem~\cite{funahashi1989approximate}, an NN is a great fit for approximating arbitrarily complex (potentially non-linear and non-convex) program behaviors.  Moreover, NNs, by design, also support efficient gradient computation that is crucial for our purposes. We train the NN by either using existing test inputs or with the test input corpus generated by existing evolutionary fuzzers as shown in Figure~\ref{fig:workflow_overview}. 



\vspace{0.2cm}
\noindent\textbf{Gradient-guided optimization.}
The smooth NN model, once trained, can be used to efficiently compute gradients and higher-order derivatives that can then be leveraged for faster convergence to the optimal solution. Different variants of gradient-guided algorithms like gradient descent, Newton's method, or quasi-Newton methods like the L-BFGS algorithm use gradients or higher-order derivatives for faster convergence~\cite{bertsekas2015convex, byrd1995limited, nocedal1980updating}.  Smooth NNs enable the fuzzing input generation process to potentially use all of these techniques. In this paper, we design, implement and evaluate a simple gradient-guided input generation scheme tailored for coverage-based fuzzing as described in detail in Section~\ref{subsec:mutation}.

\vspace{0.2cm}
\noindent\textbf{Incremental learning.} 
Any types of existing test inputs (as long as they expose diverse behaviors in the target program) can be potentially used to train the NN model and bootstrap the fuzzing input generation process. In this paper, we train the NN by collecting a set of test inputs and the corresponding edge coverage information by running evolutionary fuzzers like AFL.

However, as the initial training data used for training the NN model may only cover a small part of the program space, we further refine the model through incremental training as new program behaviors are observed during fuzzing. The key challenge in incremental training is that if an NN is only trained on new data, it might completely forget the rules it learned from old data~\cite{forgetting}. We avoid this problem by designing a new coverage-based filtration scheme that creates a condensed summary of both old and new data, allowing the NN to be trained efficiently on them.

\noindent
{\bf A Motivating Example.} We show a simple motivating example in~\Cref{fig:workflow} to demonstrate the key insight behind our approach. The simple {\tt C} code snippet shown in Figure~\ref{fig:workflow} demonstrates a general switch-like code pattern commonly found in many real-world programs.
In particular, the example code computes a non-linear exponential function of the input (\ie, \texttt{pow(3,a+b)}). It returns different values based on the output range of the computed function. Let us also assume that a buggy code block (marked in \Red{red}) is exercised if the function output range is in (1,2).  

Consider the case where evolutionary fuzzers like \afl have managed to explore the branches in lines $2$ and $9$ but fail to explore branch in line $5$. The key challenge here is to find values of \texttt{a} and \texttt{b} that will trigger the branch at line $5$. Evolutionary fuzzers often struggle with such code as the odds of finding a solution through random mutation are very low.  
For example, Figure~\ref{subfig:unsmooth} shows the original function that the code snippet represents. There is a sharp jump in the function surface from $a+b=0$ to $a+b-\epsilon=0$ ($\epsilon\rightarrow +0$). To maximize the edge coverage during fuzzing, an evolutionary fuzzer can only resort to random mutations to the input as such techniques do not consider the shape of function surface. By contrast, our NN smoothing and gradient-guided mutations are designed to exploit the function surface shape as measured by the gradients. 

We train an NN model on the program behaviors from the other two branches. The NN model smoothly approximates the program behaviors as shown in Figure~\ref{subfig:smoother} and~\ref{subfig:smoothest}.
We then use the NN model to perform more effective gradient-guided optimization to find the desired values of $a$ and $b$ and incrementally refine the model until the desired branch is found that exercises the target bug.

\section{methodology}
\label{sec:methodology}


We describe the different components of our scheme in detail below.

\subsection{Program smoothing} 
Program smoothing is an essential step to make gradient-guided optimization techniques suitable for fuzzing real-world programs with discrete behavior. Without smoothing,  gradient-guided optimization techniques are not very effective for optimizing non-smooth functions as they tend to get stuck at different discontinuities~\cite{parnas1985software}. The smoothing process minimizes such irregularities and therefore makes the gradient-guided optimization significantly more effective on discontinuous functions. 

In general, the smoothing of a discontinuous function $f$ can be thought of as a convolution operation between $f$ and a smooth mask function $g$ to produce a new smooth output function as shown below. Some examples of popular smoothing masks include different Gaussian and Sigmoid functions.

\begin{equation}
f^{\prime}(x)=\int_{-\infty}^{+\infty}f(a)g(x-a)da
\end{equation} 

However, for many practical problems, the discontinuous function $f$ may not have a closed-form representation and thus analytically computing the above-mentioned integral is not possible. In such cases, the discrete version $f^{\prime}(x)=\sum_{a}f(a)g(x-a)$ is used and the convolution is computed numerically. For example, in image smoothing, often fixed-sized 2-D convolution kernels are used to perform such computation. However, in our setting, $f$ is a computer program and therefore the corresponding convolution cannot be computed analytically. 

Program smoothing techniques can be classified into two broad categories: blackbox and whitebox smoothing. The blackbox approach picks discrete samples from the input space of $f$ and computes the convolution numerically using these samples. By contrast, the whitebox approach looks into the program statements/instructions and try to summarize their effects using symbolic analysis and abstract interpretation~\cite{chaudhuri2011smoothing, chaudhuri2010smooth}. The blackbox approaches may introduce large approximation errors while whitebox approaches incur prohibitive performance overhead, which makes them infeasible for real-world programs. 

To avoid such problems, we use NNs to learn a smooth approximation of program behaviors in a graybox manner (\eg, by collecting edge coverage data) as described below.

\subsection{Neural program smoothing}
\label{nn_smoothing}
In this paper, we propose a novel approach to program smoothing by using surrogate NN models to learn and iteratively refine smooth approximations of the target program based on the observed program behaviors. The surrogate neural networks can smoothly generalize to the observed program behaviors while also accurately modeling potentially non-linear and non-convex behaviors. The neural networks, once trained, can be used for efficiently computing gradients and higher-level derivatives to guide the fuzzing input generation process as shown in Figure~\ref{fig:workflow}.  

\smallskip
\noindent
{\bf Why NNs?} 
As implied by the universal approximation theorem~\cite{funahashi1989approximate}, an NN is a great fit for approximating complex (potentially non-linear and non-convex) program behaviors. The advantages of using NNs for learning smooth program approximations are as follows: (i) NNs can accurately model complex non-linear program behaviors and can be trained efficiently. Prior works on model-based optimization have used simple linear and quadratic models~\cite{conn1997recent, conn1997convergence, powell2002uobyqa, kolda2003optimization}. However, such models are not a good fit for modeling real-world software with highly non-linear and non-convex behaviors; (ii) NNs support efficient computation of their gradients and higher-order derivatives. Therefore, the gradient-guided algorithms can compute and use such information during fuzzing without any extra overhead; and (iii) NNs can generalize and learn to predict a program's behaviors for unseen inputs based on its behaviors on similar inputs. Therefore, NNs can potentially learn a smooth approximation of the entire program based on its behaviors for a small number of input samples. 

\smallskip
\noindent
{\bf NN Training.} 
While NNs can be used to model different aspects of a program's behavior, in this paper we use them specifically for modeling the target program's branching behavior (\ie, predicting control flow edges exercised by a given program input). One of the challenges in using neural nets to model branching behavior is the need to accept variably-sized input. Feedforward NNs, unlike real-world programs, typically accept fixed size input. Therefore, we set a maximum input size threshold and pad any smaller-sized inputs with null bytes during training. Note that supporting larger inputs is not a major concern as modern NNs can easily scale to millions of parameters. Therefore, for larger programs, we can simply increase the threshold size, if needed. However, we empirically find that relatively modest threshold values yield the best results and larger inputs do not increase modeling accuracy significantly.

Formally, let $f:\big\{\texttt{0x00}, \texttt{0x01},...,\texttt{0xff}\big\}^m \rightarrow \big\{0,1\big\}^n$ denote the NN that takes program inputs as byte sequences with size $m$ and outputs an edge bitmap with size $n$. 
Let $\theta$ denote the trainable weight parameters of $f$. 
Given a set of training samples $(X,Y)$, where $X$ is a set of input bytes and $Y$ represents the corresponding edge coverage bitmap, the training task of the parametric function $f(x,\theta)=y$ is to obtain the parameter $\hat{\theta}$ such that $\hat{\theta} = \argmin_{\theta} \sum\limits_{x\in X, y\in Y} L(y,f(x,\theta))$
where $L(y,f(x,\theta))$ defines the loss function between the output of the NN and the ground truth label $y\in Y$ in the training set.
The training task is to find the weight parameters $\theta$ of the NN $f$ to minimize the loss, which is defined using a distance metric. In particular, we use binary cross-entropy to compute the distance between the predicted bitmap and the true coverage bitmap. In particular, let $y_i$ and $f_i(x,\theta)$ denote the $i$-th bit in the output bitmap of ground truth and $f$'s prediction, respectively. Then, the binary cross-entropy between these two is defined as:

\begin{equation*}
\label{eq:cross_entropy}
\begin{split}
	-\frac{1}{n} \sum\limits_{i=1}^n [y_i\cdot log(f_i(x,\theta) + (1-y_i)\cdot log(1-f_i(x,\theta)]
\end{split}
\end{equation*}

In this paper, we use feed-forward fully connected NNs to model the target program's branching behavior. The feed-forward architecture allows highly efficient computation of gradients and fast training ~\cite{krizhevsky2012imagenet}.  

Our smoothing technique is agnostic to the source of the training data and therefore the NN can be trained on any edge coverage data gathered from an existing input corpus. For our prototype implementation, we use input corpora generated by existing evolutionary fuzzers like \afl to train our initial model. 

\medskip
\noindent\textbf{Training data preprocessing.}
Edge coverage exercised by the training data often tends to be biased, as it only contains labels for a small section of all edges in a program. For example, some edges might always be exercised together by all inputs in the training data. This type of correlation between a set of labels is known in machine learning as multicollinearity, which often prevents the model from converging to a small loss value~\cite{multicolinearity}. To avoid such cases, we follow the common machine learning practice of dimensionality reduction by merging the edges that always appear together in the training data into one edge. Furthermore, we only consider the edges that have been activated at least once in the training data. These steps significantly reduce the number of labels to around $4,000$ from around $65,536$ on average. Note that we rerun the data preprocessing step at every iteration of incremental learning and thus some merged labels may get split as their correlation may decrease as new edge data is discovered during fuzzing.

\subsection{Gradient-guided optimization}
\label{subsec:mutation}
Different gradient-guided optimization techniques like gradient descent, Newton's method, or quasi-Newton methods like L-BFGS can use gradient or higher-order derivatives for faster convergence~\cite{bertsekas2015convex, byrd1995limited, nocedal1980updating}. Smooth NNs enable the fuzzing input generation process to potentially use any of these techniques by supporting efficient computation of gradient and higher-order derivatives. In this paper, we specifically design a simple gradient-guided search scheme that is robust to minor prediction errors to demonstrate the effectiveness of our approach. We leave the exploration of more sophisticated techniques as future work.  

Before describing our mutation strategy, which is based on the NN's gradient, we first provide a formal definition of the gradient that indicates how much each input byte should be changed to affect the output of a final layer neuron in the NN (indicating changed edge coverage in the program) $f$~\cite{simonyan2013deep}. 
Here each output neuron corresponds to a particular edge and computes a value between 0 and 1 summarizing the effect of the given input byte on a particular edge. The gradients of the output neurons of the NN $f$ w.r.t. the inputs have been extensively used for adversarial input generation~\cite{goodfellow2015explain, papernot2016limitations} and visualizing/understanding DNNs~\cite{yosinskiunderstanding, simonyan2013deep, mahendran15understanding}. Intuitively, in our setting, the goal of gradient-based guidance is to find inputs that will change the output of the final layer neurons corresponding to different edges from $0$ to $1$.

Given a parametric NN $y=f(\theta, x)$ as defined in Section~\ref{nn_smoothing}, let $y_i$ denote the output of $i$-th neuron in the final layer of $f$, which can also be written as $f_i(\theta,x)$. The gradient $G$ of $f_i(\theta, x)$ with respect to input $x$ can be defined as $G=\nabla_{x}f_i(\theta, x)={\partial y_i}/{\partial x}$.
Note that $f$'s gradient w.r.t to $\theta$ can be easily computed as the NN training process requires iteratively computing this value to update $\theta$. Therefore, $G$ can also be easily calculated by simply replacing the computation of the gradient of $\theta$ to that of $x$. Note that the dimension of the gradient $\bm{G}$ is identical to that of the input $x$, which is a byte sequence in our case. 

\begin{algorithm}
\caption{Gradient-guided mutation}
\label{alg:mutation}
\lstset{basicstyle=\ttfamily\footnotesize, breaklines=true}

\begin{tabular}{|lp{2.6in}|}\hline
\textbf{Input}:
    & \textit{seed} $\leftarrow$ initial seed \\
    & \textit{iter} $\leftarrow$ number of iterations \\
    & \textit{k} $\leftarrow$ parameter for picking top-k critical bytes for mutation\\
    & \textit{g} $\leftarrow$ computed gradient of seed\\
\hline
\end{tabular}

\begin{algorithmic}[1]
\For{$i = 1$ to $iter$}
	\State $locations \gets top(g, k_i)$
	\For{$m = 1$ to $255$}
	\For{$loc \in locations$}
        \State $v \gets seed[loc] +  m * sign(g[loc])$
        \State $v \gets clip(v, 0, 255)$
        \State $gen\_mutate(seed, loc, v)$
      \EndFor
      \For{$loc \in locations$}
        \State $v \gets seed[loc] - m * sign(g[loc])$
        \State $v \gets clip(v, 0, 255)$
        \State $gen\_mutate(seed, loc, v)$
    \EndFor
  \EndFor
\EndFor
\end{algorithmic}
\end{algorithm}

\medskip
\noindent\textbf{Gradient-guided optimization.} Algorithm~\ref{alg:mutation} shows the outline of our gradient-guided input generation process. The key idea is to identify the input bytes with highest gradient values and mutate them, as they indicate higher importance to the NN and thus have higher chances of causing major changes in the program behavior (\eg, flipping branches).  

Starting from a seed, we iteratively generate new test inputs. As shown in Algorithm~\ref{alg:mutation}, at each iteration, we first leverage the absolute value of the gradient to identify the input bytes that will cause the maximum change in the output neurons corresponding to the untaken edges. Next, we check the sign of the gradient for each of these bytes to decide the direction of the mutation (\eg, increment or decrement their values) to maximize/minimize the objective function. Conceptually, our usage of gradient sign is similar to the adversarial input generation methods introduced in~\cite{goodfellow2015explain}. 
We also bound the mutation of each byte in its legal range (0-255). Lines 6 and 10 denote the use of \texttt{clip} function to implement such bounding.

We start the input generation process with a small mutation target ($k$ in Algorithm~\ref{alg:mutation}) and exponentially grow the number of target bytes to mutate to effectively cover the large input space.

\subsection{Refinement with incremental learning}
The efficiency of the gradient-guided input generation process depends heavily on how accurately the surrogate NN can model the target program's branching behavior. To achieve higher accuracy, we incrementally refine the NN model when divergent program behaviors are observed during the fuzzing process (\ie, when the target program's behavior does not match the predicted behavior).  We use incremental learning techniques to keep the NN model updated by learning from new data when new edges are triggered.

The main challenge behind NN refinement is preventing the NN model from abruptly forgetting the information it previously learned from old data while training on new data. Such forgetting is a well-known phenomenon in deep learning literature and has been thought to be a result of the stability-plasticity dilemma~\cite{mccloskey1989catastrophic, abraham2005memory}. To avoid such forgetting issues, an NN must change the weights enough to learn new tasks but not too much as to cause it to forget previously learned representations. 

The simplest way to refine an NN is to add the new training data (\ie, program branching behaviors) together with the old data and train the model from scratch again. However, as the number of data points grows, such retraining becomes harder to scale.  Prior research has tried to solve this problem using mainly two broad approaches~\cite{hinton1987using, kirkpatrick2017overcoming, fernando2017pathnet, ren2017life, draelos2017neurogenesis, goodrich2014unsupervised, robins1995catastrophic}. The first one tries to keep separate representations for the new and old models to minimize forgetting using distributed models, regularization, or creating an ensemble out of multiple models. The second approach maintains a summary of the old data and retrains the model on new data along with the summarized old data and therefore is more efficient than complete retraining. We refer the interested readers to the survey by Kemker et al.~\cite{kemker2017measuring} for more details. 

In this paper, we used edge-coverage-based filtering to only keep the old data that triggered new branches for retraining. As new training data becomes available, we identify the ones achieving new edge coverage, put them together with the filtered old training data, and retrain the NN. Such a method effectively prevents the number of training data samples from drastically increasing over the number of retraining iterations. We find that our filtration scheme can easily support up to 50 iterations of retraining while still keeping the training time under several minutes.

\begin{table*}[h!tpb]
\caption{\textbf{\small \sys Parameter Tuning}}
\label{tab:tuning}
\centering
\parbox{.49\linewidth}{
\renewcommand{\arraystretch}{1.1}
\setlength{\tabcolsep}{15pt}
\centering
    \subfloat[\textbf{\small Edge coverage achieved by mutations generated in different iterations (Algorithm~\ref{alg:mutation} line 1). The numbers in bold indicate the highest values for each program.}] {
         \label{tab:para_k}  
  \begin{tabular}{lrrr}
    \toprule
    \multirow{2}{*}{Programs} & \multicolumn{3}{c}{Iteration $i$} \\\cmidrule{2-4}
    & 7  & 10 & 11 \\
    \midrule
    readelf -a & 1,678 & \textbf{1,800} & 1,529 \\
    libjpeg & \textbf{107} & 89 & 93\\
    libxml & 161 & \textbf{256} & 174 \\
    mupdf & \textbf{294} & 266 & 266 \\
  \bottomrule
\end{tabular}
    }
}
\quad
\parbox{.46\linewidth}{
\centering
\setlength{\tabcolsep}{8pt}
\renewcommand{\arraystretch}{1.1}
\subfloat[\textbf{\small Edge coverage comparison of 1M mutations generated by \sys on different NN models. $\textbf{n}$ denotes the number of neurons in every hidden layer.}] {
  \label{tab:cn}  
  \begin{tabular}{lrr|rr}
    \toprule
    \multirow{2}{*}{Programs} & \multicolumn{2}{c|}{1 hidden layer} & \multicolumn{2}{c}{3 hidden layers} \\\cmidrule{2-5}
    & n=4096  & n=8192 & n=4096 & n=8192  \\
    \midrule
    readelf -a & 1,800& 1,658 & 1,714 & 1,584 \\
    libjpeg & 89 & 57 & 80 & 79  \\
    libxml & 256 & 172 & 140 & 99 \\
    mupdf & 260 & 94 & 82 & 88 \\
   \bottomrule
   \end{tabular}
    }
}
\end{table*}

\section{Implementation}
\label{sec:impl}

In this section, we discuss our implementation and how we fine-tune \sys to achieve optimal performance. We have released our implementation through GitHub at {\color{blue}\url{http://github.com/dongdongshe/neuzz}}. 
All our measurements are performed on a system running Arch Linux 4.9.48 with an Nvidia GTX 1080 Ti GPU.

\noindent\textbf{NN architecture.}
Our NN model is implemented in Keras-2.1.3~\cite{keras} with Tensorflow-1.4.1~\cite{tensorflow} as a backend. 
The NN model consists of three fully-connected layers. The hidden layer uses ReLU as its activation function. We use sigmoid as the activation function for the output layer to predict whether a control flow edge is covered or not. The NN model is trained for 50 epochs (\ie, 50 complete passes of the entire dataset) to achieve high test accuracy (around $95\%$ on average). 
Since we use a simple feed-forward network, 
the training time for all 10 programs is less than 2 minutes. Even with pure CPU computation on an Intel i7-7700 running at 3.6GHz, the training time is under 20 minutes. 

\noindent\textbf{Training Data Collection.}
For each program tested, we run \afl-2.52b~\cite{afl} on a single core machine for an hour to collect training data for the NN models. 
The average number of training inputs collected for $10$ programs is around $2K$. 
The resulting corpus is further split into training and testing data with a 5:1 ratio, where the testing data is used to ensure that the models are not overfitting. We use \texttt{10KB} as the threshold file size for selecting our training data from the AFL input corpus (on average 90\% of the files generated by AFL were under the threshold).  



\noindent
\textbf{Mutation and Retraining.}
As shown in~\Cref{fig:workflow_overview}, \sys runs iteratively to generate 1M mutations and incrementally retrain the NN model.
We first use the mutation algorithm described in~\Cref{alg:mutation} to generate 1M mutations. We set the parameter \texttt{i} to 10, which generates 5,120 mutated inputs for a seed input. Next, we randomly choose 100 output neurons representing 100 unexplored edges in the target program and generate 10,240 mutated inputs from two seeds. Finally, we execute the target program with 1M mutated inputs using \afl's fork server technique~\cite{forkserver} and use any inputs covering new edges for incremental retraining. 

\noindent
\textbf{Model Parameter Selection.} 
The success of \sys depends on the choices of different parameters in training the models and generating mutations.
Here, we empirically explore the optimal parameters that ensure maximum edge coverage on four programs:  \texttt{readelf}, \texttt{libjpeg}, \texttt{libxml}, and \texttt{mupdf}. The results are summarized in Table~\ref{tab:tuning}. 

First, we evaluate how many critical bytes need to be mutated per initial seed (parameter $k_i$ in line 1 of ~\Cref{alg:mutation}). 
We choose $k=2$ as described in~\Cref{subsec:mutation} and show the coverage achieved by three iterations ($i=7,10,11$ in Algorithm~\ref{alg:mutation} line 1) with $1$M mutations per iteration. For all four programs, \emph{smaller} mutations (with fewer bytes changed per mutation) may lead to higher code coverage, as shown in Table~\ref{tab:para_k}. 
The largest value of $i=11$ achieves the least code coverage for all four programs. This result is potentially due to lines 4 and 8 in~\Cref{alg:mutation}\textemdash wasting too many mutations  (out of the 1M mutation budget) on a single seed, without trying other seeds. 
However, the optimal number of mutation bytes varies across the four programs. For \texttt{readelf} and \texttt{libxml}, the optimal value of $i$ is $10$, while it is $7$ for  \texttt{libjpeg} and \texttt{mupdf}. Since the difference in achieved code coverage between $i=7$ and $i=10$ is not large, we choose $i=10$ for the remainder of the experiments.  

Next, we evaluate the choice of hyper-parameters in the NN model by varying the number of layers and the number of neurons in each hidden layer.  In particular, we compare NN architectures with $1$ and $3$ hidden layers and $4096$ and $8192$ neurons per layer, respectively. For every target program, we use the same training data to train four different NN models and generate 1M mutations to test the achieved edge coverage. For all four programs, we find that the model with $1$ hidden layer performs better than the one with $3$ hidden layers. We think this is because the 1 hidden layer model is sufficiently complex to model the branching behavior of the target program, whereas the larger model (\ie, with 3 hidden layers) is relatively harder to train and also tends to overfit.

\begin{table*}[b!htp]
\caption{\textbf{Study Subjects}}
\label{tab:subj}
\centering
\parbox{.46\linewidth}{
\renewcommand{\arraystretch}{1.25}
\setlength{\tabcolsep}{1pt}
\centering
    \subfloat[\textbf{\small Studied Fuzzers}] {
    \label{subtab:fuzzer}  
        \begin{tabular}[t]{lp{0.7\columnwidth}}
        \toprule
        \textbf{Fuzzer} & \textbf{Technical Description} \\
        \toprule
        \afl~\cite{afl} & evolutionary search \\
        \aflfast~\cite{aflfast} & evolutionary + markov-model-based search\\
        Driller~\cite{driller}$^\ddagger$ & evolutionary + concolic execution\\
        \vuzzer~\cite{vuzzer} & evolutionary + dynamic-taint-guided search \\
        \kleefl~\cite{kleefl} & evolutionary + seeds generated by symbolic execution \\
        \lafintel~\cite{lafintel} & evolutionary + transformed compare instruction\\
        RNNfuzzer~\cite{rajpal2017not} & evolutionary + RNN-guided mutation filter\\
        Steelix~\cite{steelix}$^\dag$ & evolutionary + instrumented comparison instruction\\
        T-fuzz~\cite{tfuzz}$^\dag$ & evolutionary + program transformation\\
        Angora~\cite{angora}$^\dag$ & evolutionary + dynamic-taint-guided + coordinate descent + type inference \\
        \bottomrule
        \multicolumn{2}{p{0.46\textwidth}}{$^\dag$ We only compare based on the reported LAVA-M results as they are either not open-source or do not scale to our test programs.} \\
        \multicolumn{2}{p{0.46\textwidth}}{$^\ddagger$ We only compare based on CGC as Driller only supports CGC binaries.} \\
        \end{tabular}
    }
}
\hfill
\parbox{.46\linewidth}{
\centering
\setlength{\tabcolsep}{3pt}
\renewcommand{\arraystretch}{1.1}
\subfloat[\textbf{\small Studied Programs}] {
\label{subtab:prog}
\begin{tabular}[t]{ll|r|c|c}
\toprule
\multicolumn{2}{c|}{\bf Programs} & \multirow{2}{*}{\bf \# Lines~} & \multicolumn{1}{c|}{\multirow{2}{*}{\begin{tabular}[c]{@{}c@{}}\textbf{\sys}\\ \textbf{train (s)}\end{tabular}}} & \multicolumn{1}{c}{\bf AFL coverage} \\ \cmidrule{1-2} 
Class & Name &  & \multicolumn{1}{r|}{} & 1 hour  \\ \midrule
\multirow{5}{*}{\begin{tabular}[l]{@{}l@{}}binutils-2.30~~\\ ELF\\ Parser\end{tabular}} 
 & readelf -a &21,647  & 108 & 4,490  \\
 & nm -C & 53,457 & 63 & 3,779 \\
 & objdump -D& 72,955 & 104 & 5,196 \\
 & size & 52,991 & 52 & 2,578  \\
 & strip & 56,330 & 55 & 5,789  \\ \midrule
TTF & harfbuzz-1.7.6 & 9,853 & 94 & 82,79  \\ \midrule
JPEG & libjpeg-9c & 8,857 & 56 & 3,117  \\ \midrule
PDF & mupdf-1.12.0 & 123,562 & 62 & 4,624  \\ \midrule
XML & libxml2-2.9.7 & 73,920 & 95 & 6,691  \\ \midrule
Zip & zlib-1.2.11 & 1,893 & 65 & 1,479 \\ \bottomrule
\end{tabular}
    }
}
\vspace{-0.2cm}
\end{table*}

\section{Evaluation}
\label{sec:eval}

In this section, we evaluate \sys's bug finding performance and achieved edge coverage with respect to other state-of-the-art fuzzers. Specifically, we answer the following four research questions:

\begin{itemize}[leftmargin=*]
    \item
    \textbf{RQ1.} \rqa
    \item
    \textbf{RQ2.} \rqb
    \item
    \textbf{RQ3.} \rqc
    \item
    \textbf{RQ4.} \rqd
\end{itemize}

We start by describing our study subjects and experimental setting.

\subsection{Study Subjects}
We evaluate \sys on three different types of datasets: 
(i) 10 real-world programs, as shown in~\Cref{subtab:prog}, 
(ii) LAVA-M~\cite{lava}, and 
(iii) the DARPA CGC dataset~\cite{cgc_repo}. 
To demonstrate the performance of \sys, we compare the edge coverage and number of bugs detected by \sys to $10$ state-of-the-art fuzzers, as shown in~\Cref{subtab:fuzzer}.

\subsection{Experimental Setup}

Our experimental setup includes the following two steps:
First, we run \afl for an hour to generate the initial seed corpus. 
Then, we run each fuzzer for a fixed time budget with the same initial seed corpus and compare their achieved edge coverage and the number of bugs found. Specifically, the time budgets for 10 real world programs, LAVA-M datasets and CGC datasets are 24 hours, 5 hours, and 6 hours respectively. For evolutionary fuzzers, the seed corpus is used to initialize the fuzzing process. For learning-based fuzzers (\ie, \sys and RNN-based fuzzers), the same seed corpus is used to generate the training dataset. As for \kleefl, a hybrid tool consisting of Klee and \afl, we run Klee for an extra hour to generate additional seeds, then add them into the original seed corpus for the following 24 hour fuzzing process. Note that we only report the additional code covered by the mutated inputs of each fuzzer without including the coverage information from the initial seed corpus. 
    

In RQ\ref{req:rnn}, we evaluate and compare the performance of \sys with that of the RNN-based fuzzers. The RNN-based fuzzers could take up to 20$\times$ longer training time than \sys. However, to focus on the efficacy of these two mutation algorithms, we evaluate the edge coverage for a fixed amount of mutations to exclude the effect of these disparate training time. We also perform a standalone evaluation comparing the training time costs for these two models. In RQ\ref{req:model}, we also evaluate the edge coverage for a fixed number of mutations to exclude the effect of varying training time cost across different models.  





\subsection{Results}
\label{subsec:result}

\RQA{1}{\rqa}

To answer this RQ, we evaluate \sys \wrt other fuzzers in three settings: 
(i) Detecting real-world bugs.
(ii) Detecting injected bugs in LAVA-M dataset~\cite{lava}.
(iii) Detecting CGC bugs.
We describe the results in details.

\medskip
\noindent
\textbf{(i) Detecting real-world bugs.}
We compare the total number of bugs and crashes found by \sys and other fuzzers on 24-hour running time given the same seed corpus.
There are five different types of bugs found by \sys and other fuzzers: out-of-memory, memory leak, assertion crash, integer overflow, and heap overflow. To detect memory bugs that would not necessarily lead to a crash, we compile program binaries with AddressSanitizer~\cite{asan}. We measure the unique memory bugs found by comparing the stack traces reported by AddressSanitizer. For crashes that do not cause AddressSanitizer to generate a bug report, we examine the execution trace. The integer overflow bugs are found by manually analyzing the inputs that trigger an infinite loop. 
We further verify integer overflow bugs using undefined behavior sanitizer~\cite{usan}.
The results are summarized in Table~\ref{tab:bug_summary}. 

\sys finds all $5$ types of bugs across $6$ programs. \afl, \aflfast, and \lafintel find $3$ types of bugs\textemdash they do not find any integer overflow bugs.  
The other fuzzers only uncover $2$ types of bugs (\ie, memory leak and assertion crash). \afl can a heap overflow bug on program \texttt{size}, while \sys can find the same bug and another heap overflow bug on program \texttt{nm}. In total, \sys finds $2\times$ more bugs than the second best fuzzer. Moreover, the integer-overflow bug in \texttt{strip} and the heap-overflow bug in \texttt{nm}, \textit{only found by \sys}, have been assigned with CVE-2018-19932 and CVE-2018-19931, later fixed by the developers . 
\begin{table}[h!tpb]
\setlength{\tabcolsep}{2pt}
\newcommand{\cmark}{\ding{51}}%
\newcommand{\xmark}{\ding{55}}%
\centering
\renewcommand{\arraystretch}{1.1}
\caption{\textbf{\small Number of real-world bugs found by $\mathbf{6}$ fuzzers. 
We only list the programs where the fuzzers find a bug.}}
\label{tab:bug_summary}
\begin{tabular}{lcccccc}
\toprule
Programs & \afl & \aflfast & \vuzzer & \kleefl & \lafintel & \sys \\ \midrule
\multicolumn{7}{c}{Detected Bugs per Project} \\ \midrule
readelf & 4 & 5 & 5 & 3 & 4 & 16\\
nm & 8 & 7 & 0 & 0 & 6 & 9 \\
objdump & 6 & 6 & 0 & 3 & 7 & 8\\
size & 4 & 4 & 0 & 3 & 2 & 6\\
strip & 7 & 5 & 2 & 5 & 7 & 20\\
libjpeg & 0 & 0 & 0 & 0 & 0 & 1\\
\midrule
\multicolumn{7}{c}{Detected Bugs per Type} \\ \midrule
out-of-memory & \cmark& \cmark & \xmark & \cmark & \cmark & \cmark \\
memory leak & \cmark & \cmark & \cmark & \cmark & \cmark & \cmark \\
assertion crash & \xmark & \cmark & \xmark & \xmark &  \cmark & \cmark\\
interger overflow & \xmark & \xmark & \xmark & \xmark & \xmark & \cmark \\
heap overflow & \cmark& \xmark & \xmark & \xmark & \xmark & \cmark \\

\midrule
Total & 29 & 27 & 7 & 14 & 26 & 60 \\
\bottomrule
\end{tabular}
\end{table}


\medskip
\noindent
\textbf{(ii)~Detecting injected bugs in LAVA-M dataset.} 
The LAVA dataset is created to evaluate the efficacy of fuzzers by providing a set of real-world programs injected with a large number of bugs~\cite{lava}. LAVA-M is a subset of the LAVA dataset, consisting of $4$ GNU coreutil programs {\tt base64}, {\tt md5sum}, {\tt uniq}, and {\tt who} injected with $44$, $57$, $28$, and $2136$ bugs, respectively. All the bugs are guarded by four-byte magic number comparisons. The bugs get triggered only if the condition is satisfied. We compare \sys's performance at finding these bugs to other state-of-the-art fuzzers, as shown in~\Cref{tab:lava}. 
Following conventional practice~\cite{angora,lava}, we use 5-hour time budget for the fuzzers' runtime. 

Triggering a magic number condition in the LAVA dataset is a hard task for a coverage-guided fuzzer because the fuzzer has to generate the exact combination of 4 continuous bytes out of $256^4$ possible cases. To solve this problem, we used a customized LLVM pass to instrument the magic byte checks like Steelix~\cite{steelix}. But unlike Steelix, we leverage the NN's gradient to guide the input generation process to find an input that satisfies the magic check. 
We run \afl for an hour to generate the training data and use it to train an NN whose gradients identify the possible critical bytes triggering the first byte-comparison of a magic-byte condition. Next, we perform a locally exhaustive search on each byte adjacent to the first critical byte to solve each of the remaining three byte-comparisons with $256$ tries. Therefore, we need one NN gradient computation to find the byte locations that affect the magic checking and $4\times256 = 1024$ trials to trigger each bug. 
For program {\tt md5sum}, following the latest suggestion of the LAVA-M's authors~\cite{suggestion}, we further reduce the seed into a single line, which significantly boosts the fuzzing performance. 

As shown in Table~\ref{tab:lava},
\sys finds all the bugs in programs {\tt base64}, {\tt md5sum}, and {\tt uniq}, and the highest number of bugs for program {\tt who}.  
Note that LAVA-M authors left some bugs unlisted in all $4$ programs, so the total number of bugs found by \sys is actually higher than the number of listed bugs, as shown in the result.

\sys has two key advantages over the other fuzzers. First, \sys breaks the search space into multiple manageable steps: \sys trains the underlying NN on \afl generated data, uses the computed gradient to reach the first critical byte, and performs a local search around the found critical region.
Second, as opposed to \vuzzer, which leverages magic numbers hard-coded in the target binary to construct program inputs, \sys's gradient-based searching strategy do not rely on any hard-coded magic number. Thus, it can find all the bugs in program {\tt md5sum}, which performs some computations on the input bytes before the magic number checking causing \vuzzer to fail. In comparison to Angora, the current state-of-the-art fuzzer for LAVA-M dataset, \sys finds $3$ more bugs in {\tt md5sum}. Unlike Angora, \sys uses NN gradients to trigger the complex magic number conditions more efficiently. 

\begin{table}[h!tpb]
\vspace{0.2cm}
\setlength{\tabcolsep}{8 pt}
\centering
  \caption{\textbf{\small Bugs found by different fuzzers on LAVA-M datasets.}}
  \label{tab:lava}  
\begin{tabular}{lrrrr}
\toprule
         &	base64	& md5sum & uniq & who \\ \midrule
 \#Bugs  &	44	& 57     &  28	& 2,136 \\
 FUZZER  & 	7	& 2	     & 7	& 0 \\
 SES 	 &  9	& 0	& 0	& 18 \\
 \vuzzer & 	17	& 1	& 27	& 50 \\
 Steelix &	43	& 28	& 24	& 194 \\
 Angora  &	48	& 57	& 29	& 1,541 \\
 \lafintel 	 &  42	& 49	& 24	& 17 \\
 T-fuzz  &	43	& 49	& 26	&63 \\
 \sys 	 &  48	& 60	& 29	&1,582 \\
 \bottomrule
\end{tabular}
\end{table}

\medskip
\noindent
\textbf{(iii) Detecting CGC bugs.}
The DARPA CGC dataset~\cite{cgc_corpus} consists of vulnerable programs used in the DARPA Cyber Grand Challenge. These programs are implemented as network services performing various tasks and aim to mirror real-world applications with known vulnerabilities. Every bug in the program is guarded by a number of sanity checks on the input. The dataset comes with a set of inputs as proof of vulnerabilities.

We evaluate \sys, Driller, and \afl on 50 randomly chosen CGC binaries. As running each test binary for each fuzzer takes 6 hours to run on CPU/GPU and our limited GPU resources do not allow us to execute multiple instances in parallel, we randomly picked 50 programs to keep the total experiment time within reasonable bounds. Similar to LAVA-M, here we also run  AFL  for an hour to generate the training data and use it to train the NN. We provide the same random seed to all three fuzzers and let them run for six hours. \sys uses the same customized LLVM pass used for the LAVA-M dataset to instrument magic checkings in CGC binaries. 


\begin{table}[!t]
\setlength{\tabcolsep}{5pt}
\centering
\renewcommand{\arraystretch}{1.5}
	\caption{\textbf{\small Bugs found by 3 fuzzers in 50 CGC binaries}}
  \begin{tabular}{l|rrr}
    \toprule
    Fuzzers & \afl &  Driller & \sys\\
    Bugs  & 21 & 25 & 31 \\
  \bottomrule
\end{tabular}
\label{cgc_results}
\end{table}

The results (Table~\ref{cgc_results}) show that \sys uncovers 31 buggy binaries out of 50 binaries, while \afl and Driller find 21 and 25, respectively. The buggy binaries found by \sys include all those found by Driller and \afl.
\sys further found bugs in 6 new binaries that both \afl and Driller fail to detect.
\lstset{basicstyle=\footnotesize\ttfamily,breaklines=true}
\label{cgc_code}
\begin{lstlisting}[caption={\textbf{\small \texttt{cgc\_ReceiveCommand} function in CROMU\_00027}}, captionpos=b, label={lst:run_ex}, float=h]
int cgc_ReceiveCommand(CommandStruct* command, 
  int* more_command){
  ...
  if(cgc_strncmp(&buffer[1], "VISUALIZE",
    cgc_strlen("VISUALIZE")) == 0){
    command->command = VISUALIZE; 
    //vulnerable code
  ...
\end{lstlisting}
\vspace{2pt}

We analyze an example program \texttt{CROMU\_00027} (shown in Listing~\ref{cgc_code}). This is an ASCII content server that takes a query from a client and serves the corresponding ASCII code. 
A null-pointer dereferencing bug is triggered after a user tries to set command as VISUALIZE. \afl failed to detect this bug within 6-hour time budget due to its inefficiency at guessing the magic string. Although Driller tries to satisfy such complex magic string checking by concolic execution, in this case it fails to find an input that satisfies the check. By contrast, \sys can easily use the NN gradient to locate the critical bytes in the program input that affects the magic comparison and find inputs that satisfy the magic check.

\RS{1}{\sys found 31 previously unknown bugs in 6 different programs that other fuzzers could not find. \sys also outperforms the state-of-the-art fuzzers at finding LAVA-M and CGC bugs.
}

\begin{figure*}
\centering
\captionsetup[subfloat]{captionskip=-.01cm, labelformat=empty}

\includegraphics[scale=0.8]{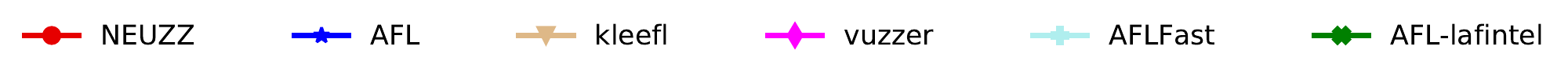}

\subfloat[readelf]{
\includegraphics[width=0.18\textwidth]{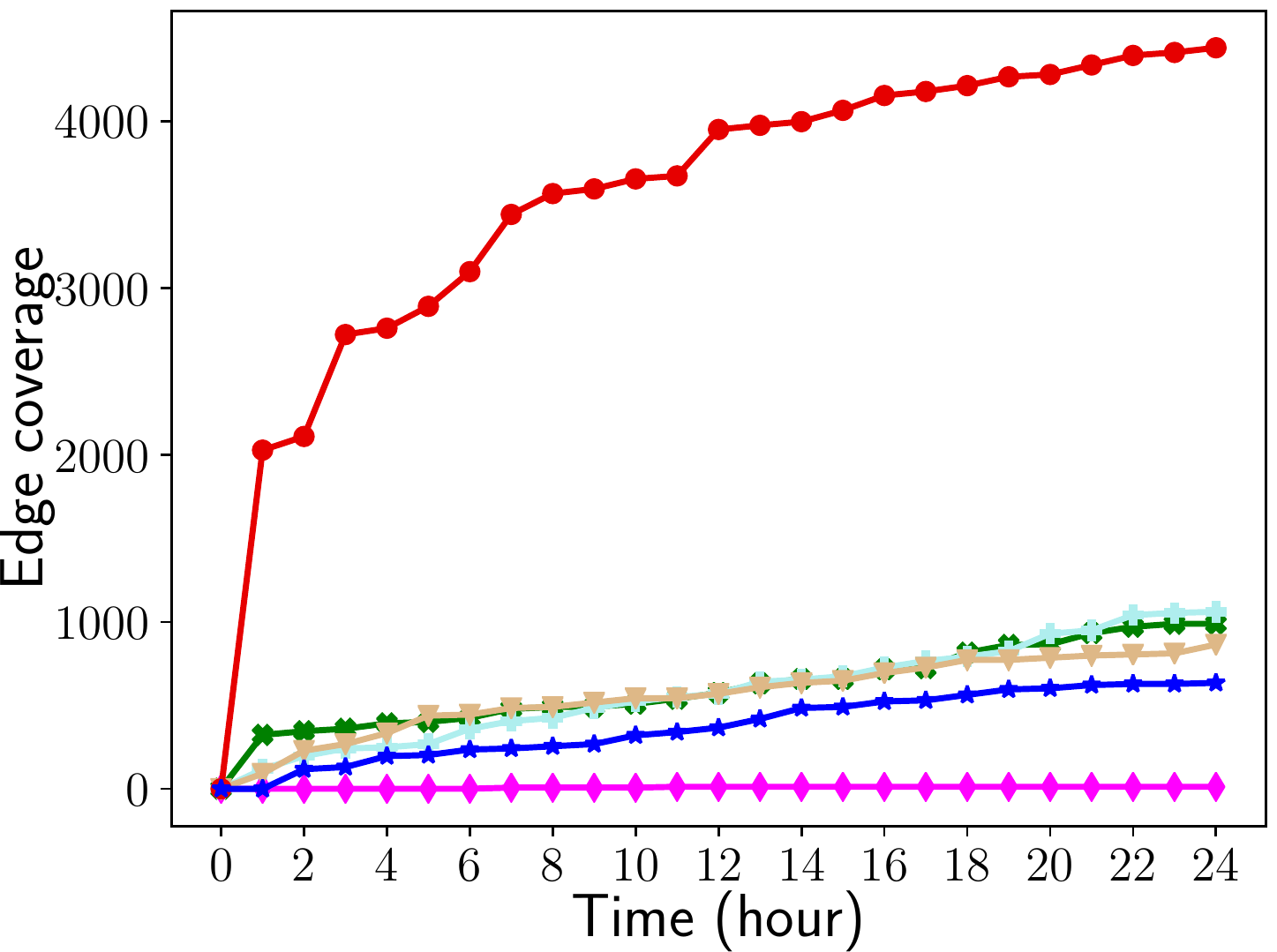}
\label{subfig:readelf_1h}}
\hspace{-.1cm}
\subfloat[harfbuzz]{
\includegraphics[width=0.18\textwidth]{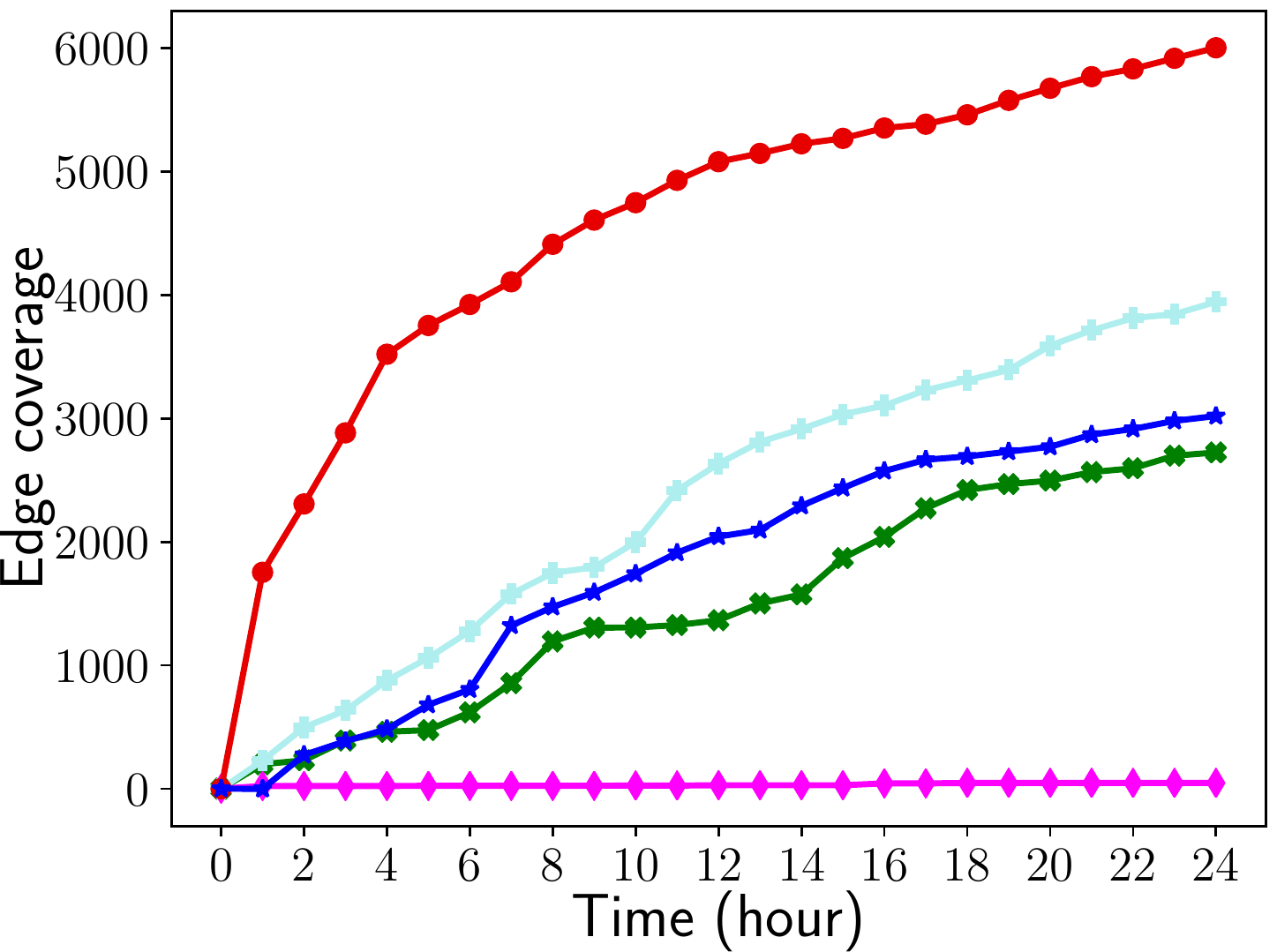}
\label{subfig:harfbuzz_1h}}
\hspace{-.1cm}
\subfloat[libjpeg]{
\includegraphics[width=0.18\textwidth]{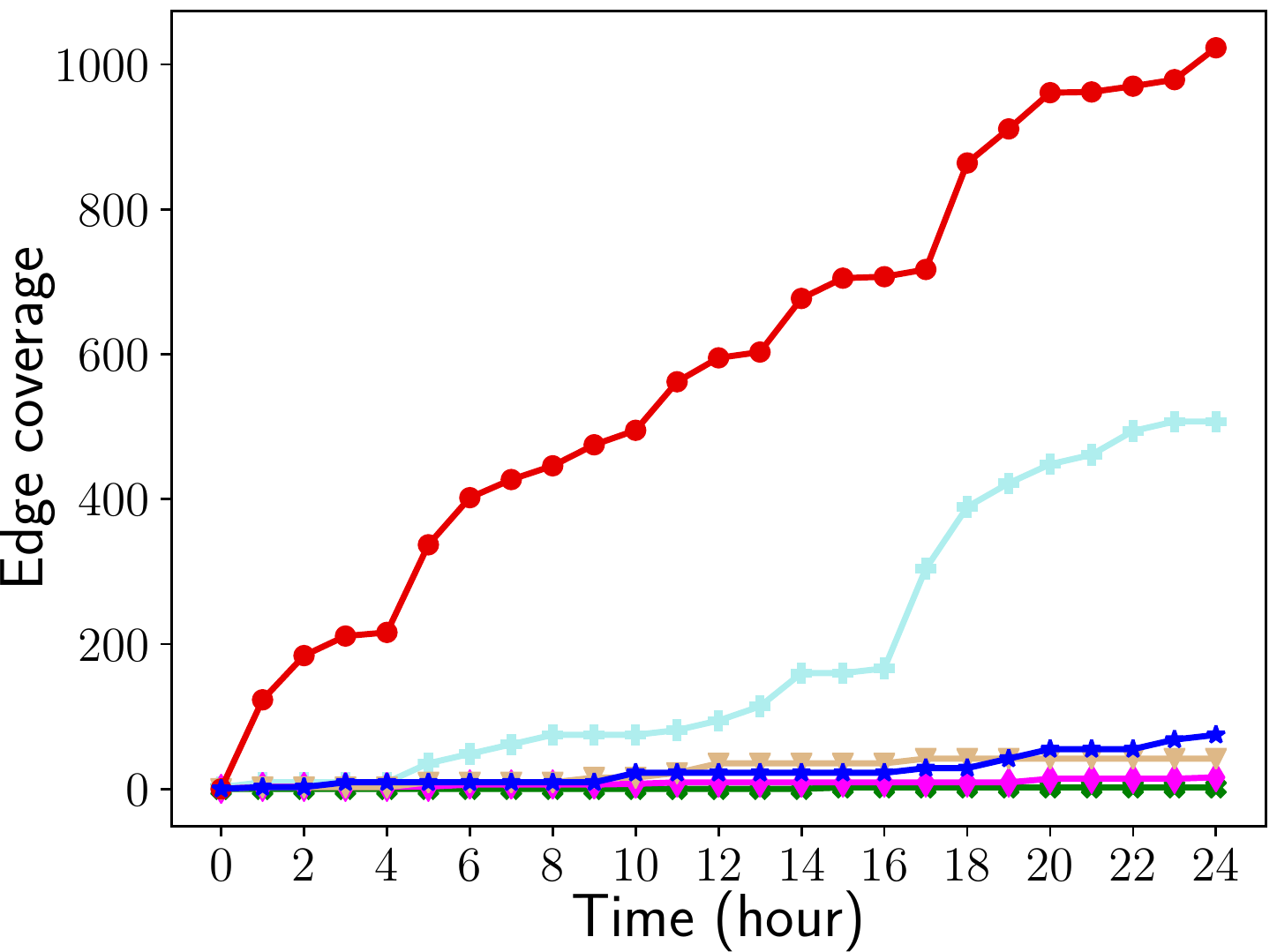}
\label{subfig:libjpeg_1h}}
\hspace{-.1cm}
\subfloat[mupdf]{
\includegraphics[width=0.18\textwidth]{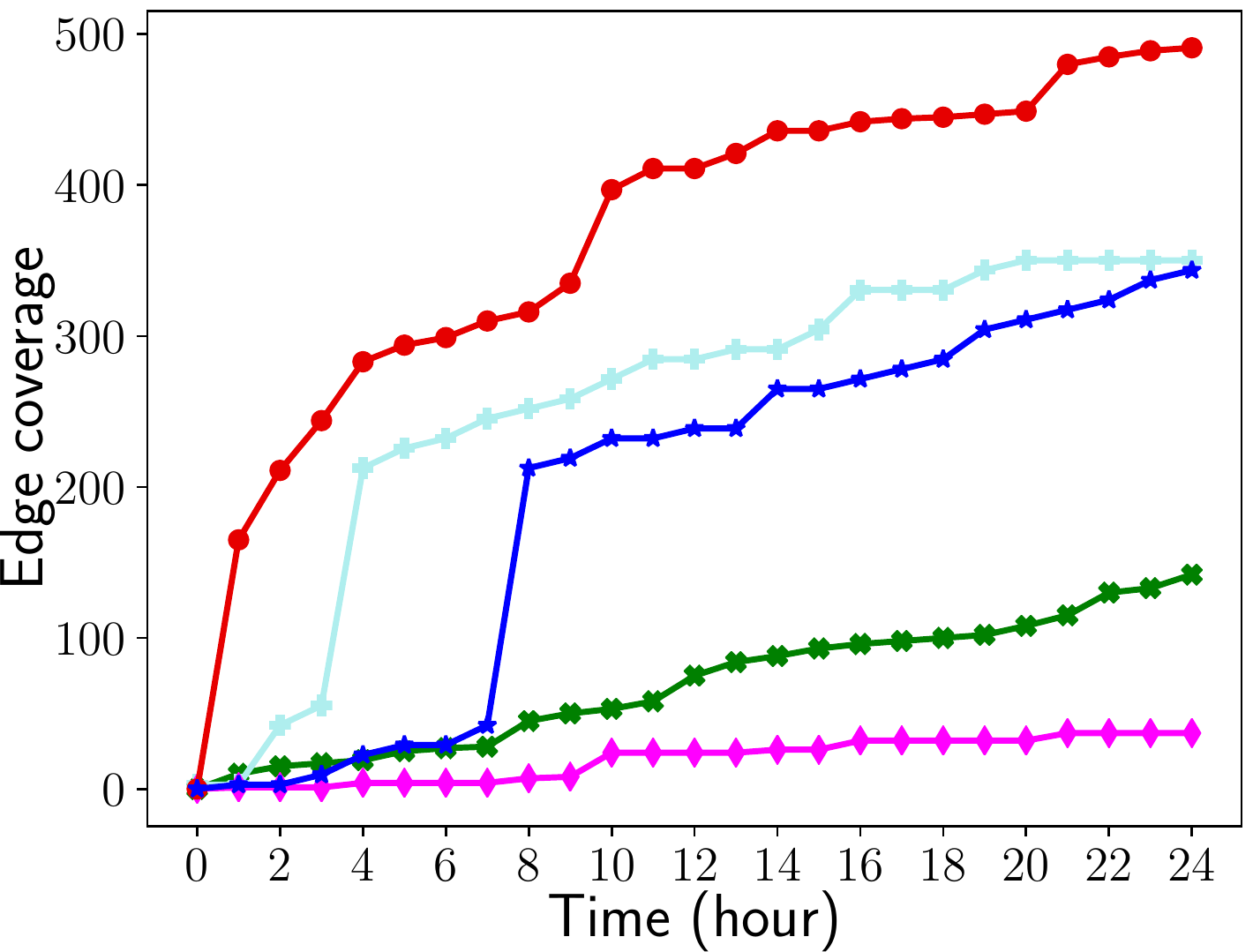}
\label{subfig:mupdf_1h}}
\hspace{-.1cm}
\subfloat[libxml]{
\includegraphics[width=0.18\textwidth]{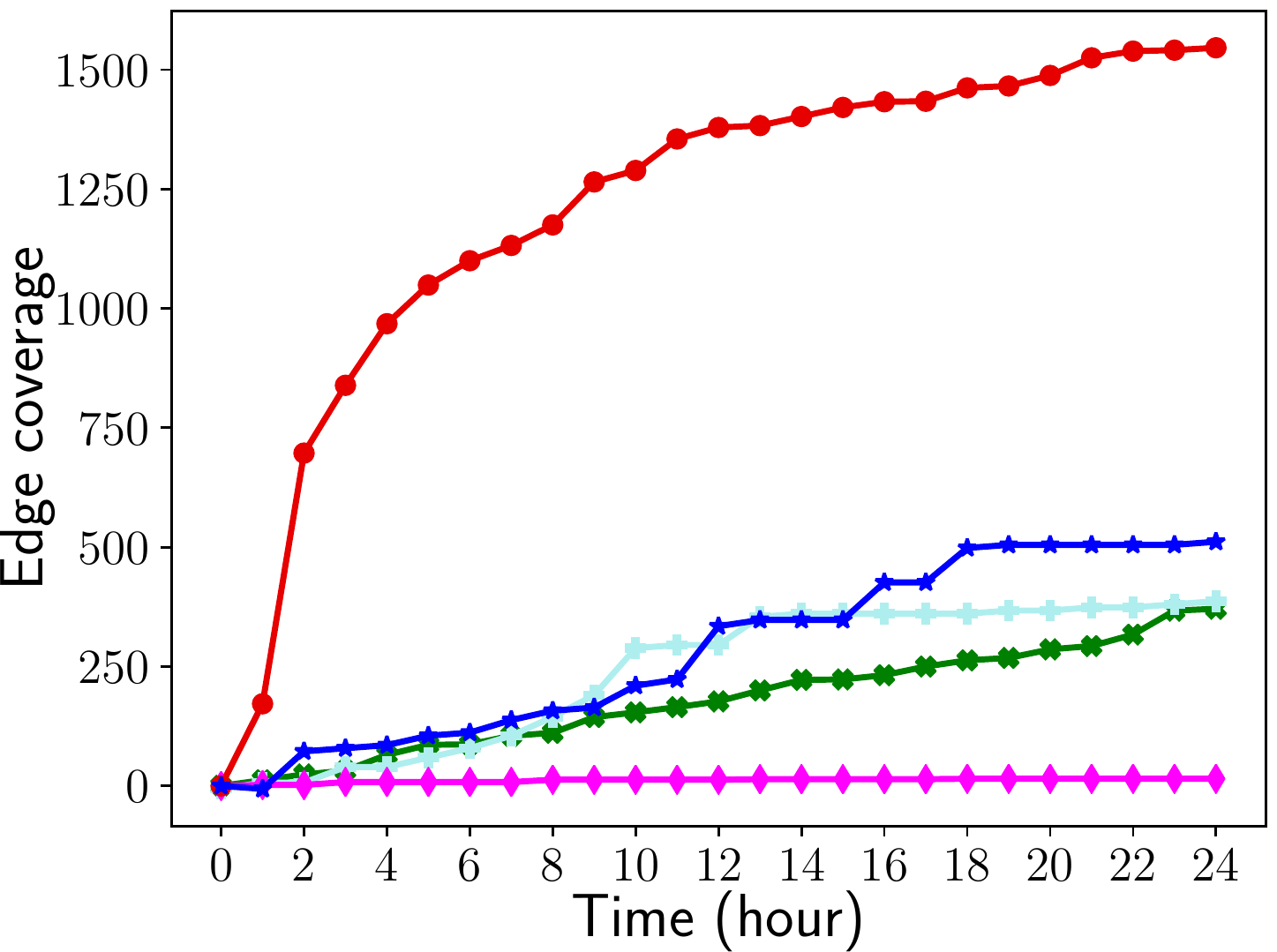}
\label{subfig:libxml_1h}}

\subfloat[nm]{
\includegraphics[width=0.18\textwidth]{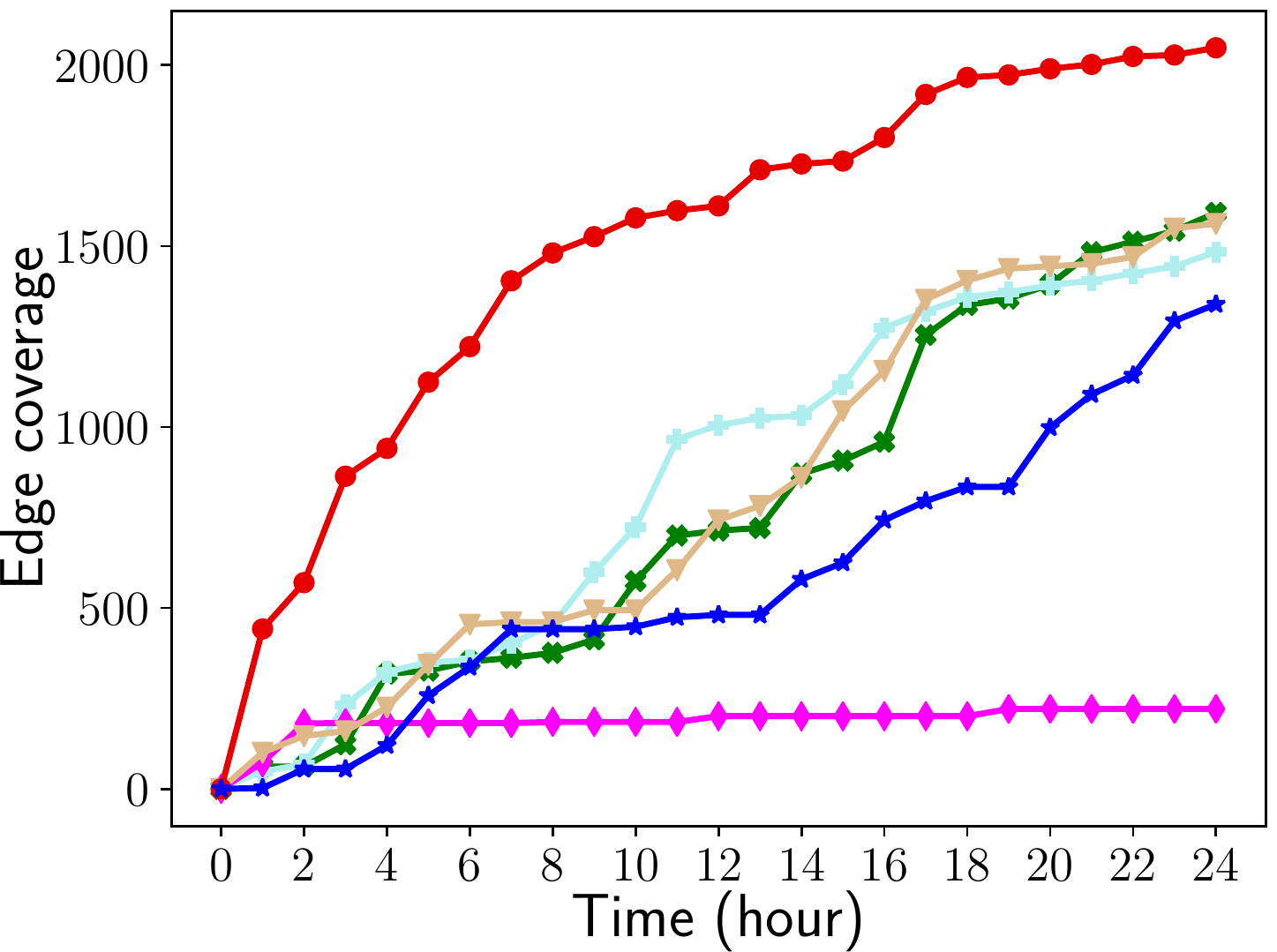}
\label{subfig:nm_1h}}
\hspace{-.1cm}
\subfloat[objdump]{
\includegraphics[width=0.18\textwidth]{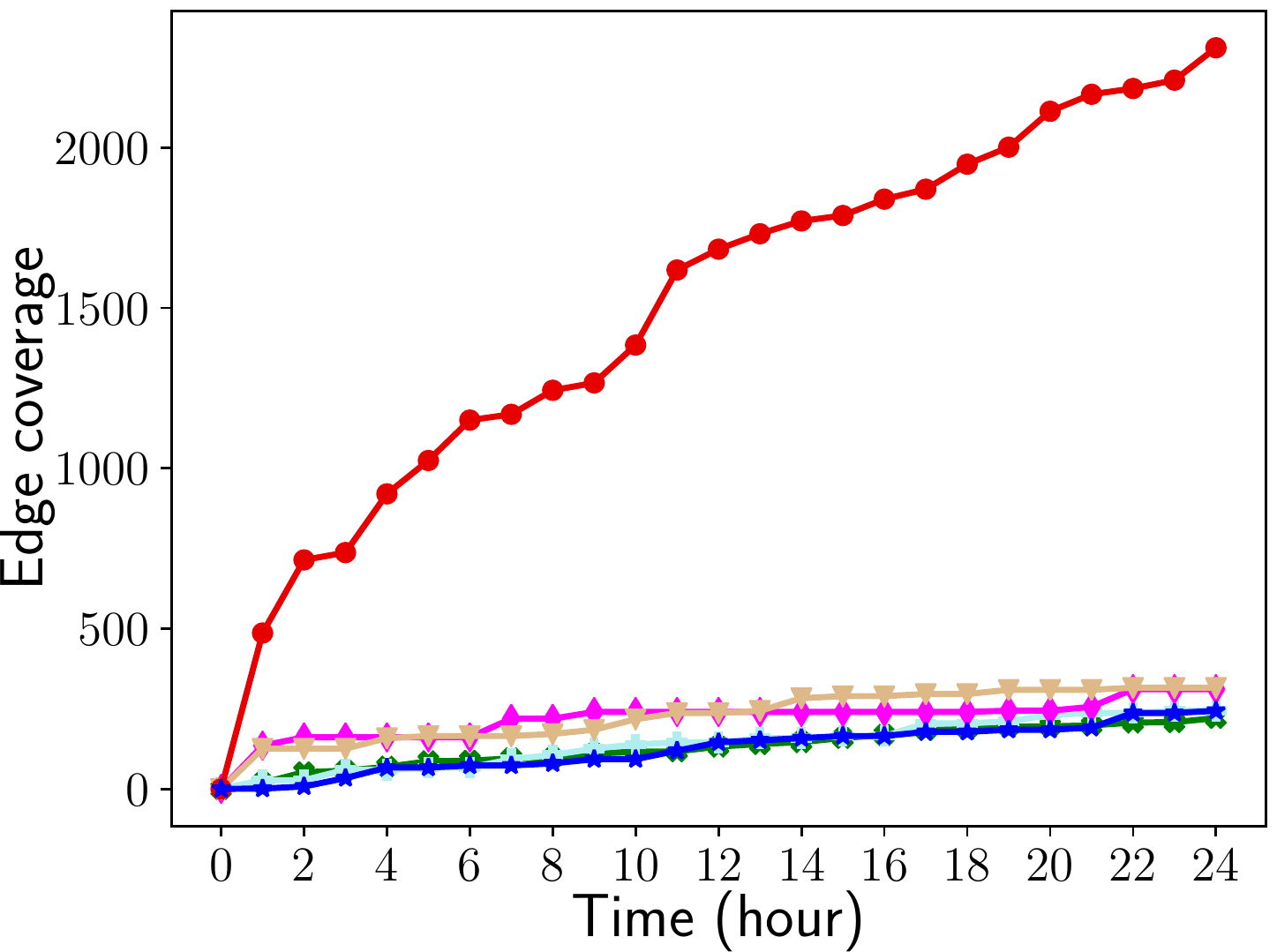}
\label{subfig:objdump_1h}}
\hspace{-.1cm}
\subfloat[size]{
\includegraphics[width=0.18\textwidth]{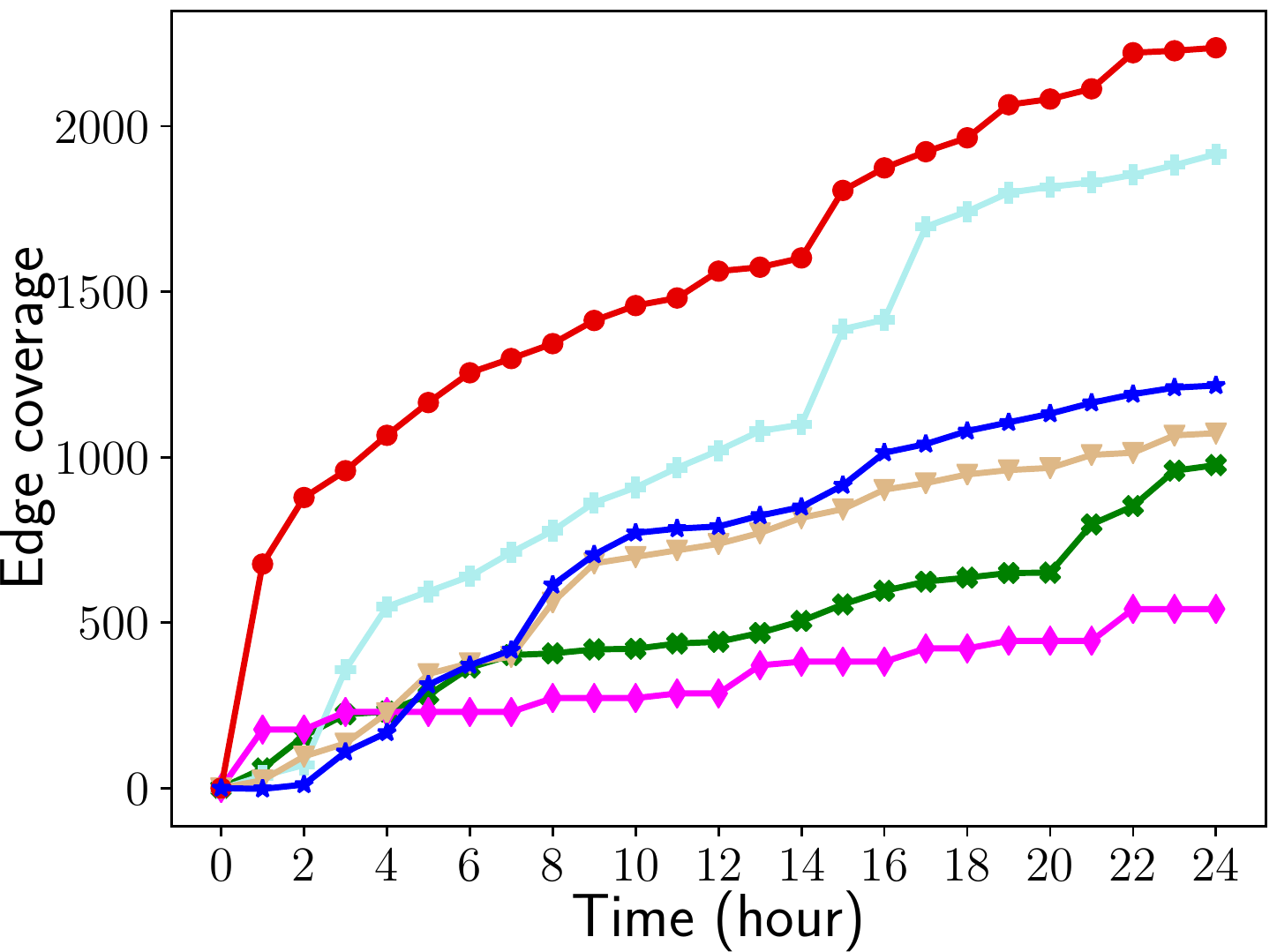}
\label{subfig:size_1h}}
\vspace{-.1cm}
\subfloat[strip]{
\includegraphics[width=0.18\textwidth]{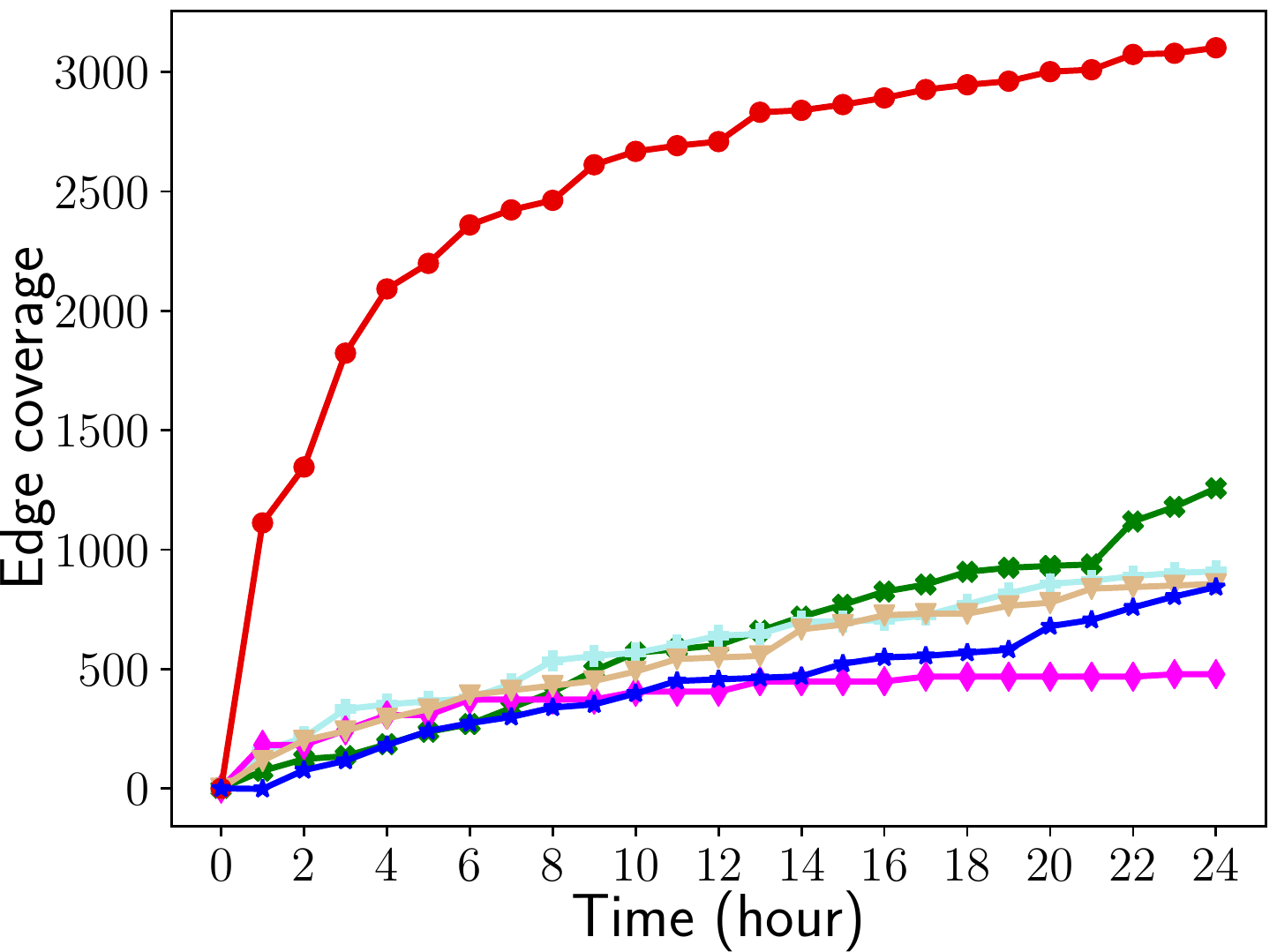}
\label{subfig:strip_1h}}
\vspace{-.1cm}
\subfloat[zlib]{
\includegraphics[width=0.18\textwidth]{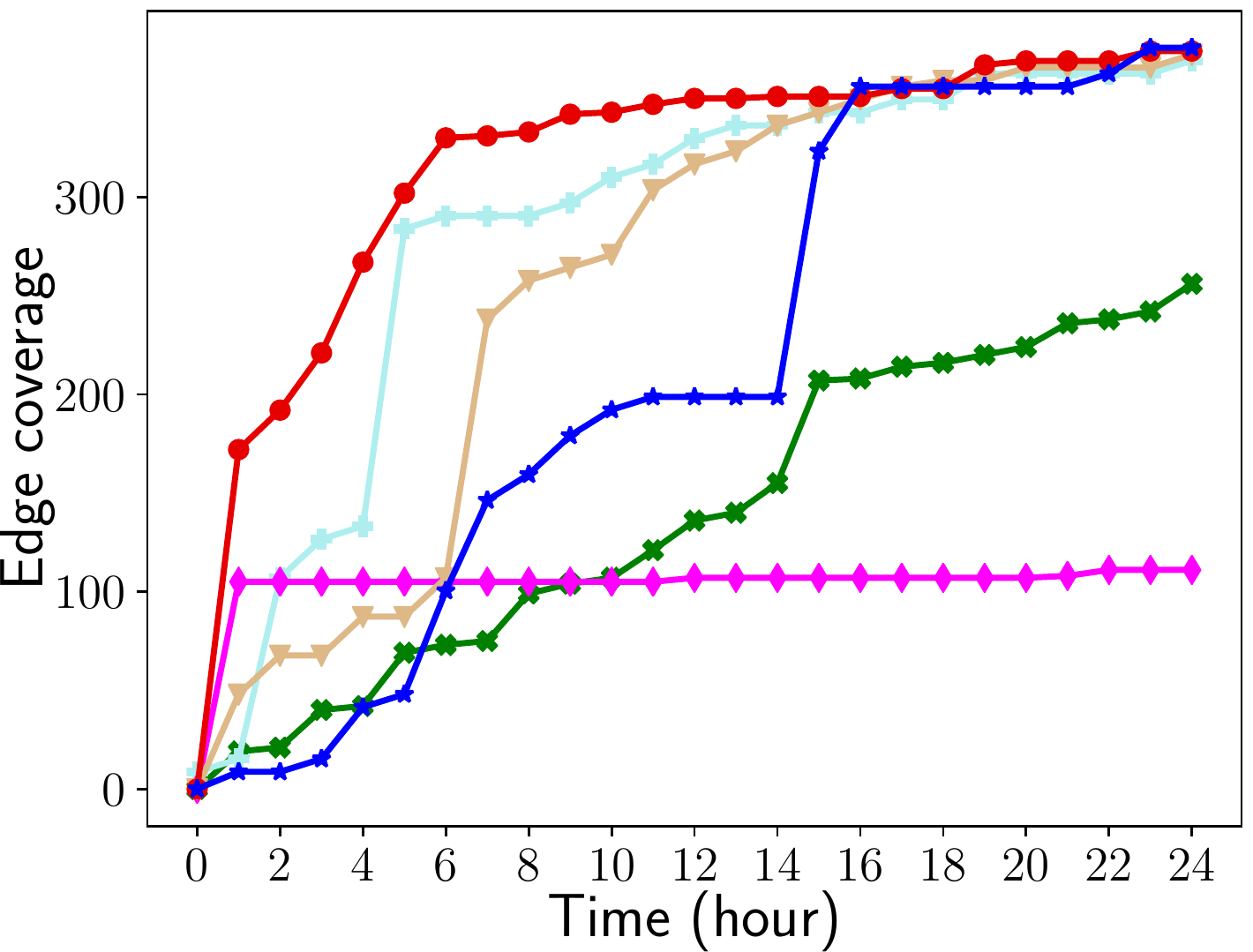}
\label{subfig:zlib_1h}}

\caption{\textbf{\small The edge coverage of different fuzzers running for 24 hours.}}
\label{fig:nn_vs_AFL}
\end{figure*}

\RQA{2}{\rqb}

To investigate this question, we compare the fuzzers on 24-hour fixed runtime budget. This evaluation shows not only the total number of new edges found by fuzzers but also the speed of new edge coverage versus time. 

\begin{table}[!htpb]
\setlength{\tabcolsep}{2pt}
\renewcommand{\arraystretch}{1.1}
\centering
\caption{\textbf{\small Comparing edge coverage of \sys \wrt other fuzzers for 24 hours runs.}}
\label{tab:cov}
\begin{tabular}{lrrrrrr}
    \toprule
   Programs & \textbf{\sys}  & \textbf{\afl} & \textbf{\aflfast} & \textbf{\vuzzer} & \textbf{\kleefl} & \textbf{\lafintel}\\ 
    \midrule
    readelf -a & 4,942 & 746 & 1,073 & 12 & 968 & 1,023\\
    nm -C & 2,056 & 1,418 & 1,503 & 221 & 1,614 & 1,445\\
    objdump -D & 2,318 & 257 & 263 & 307 & 328 & 221\\
    size & 2,262 & 1,236 & 1,924 & 541 & 1,091 & 976 \\
    strip & 3,177 & 856 & 960 & 478 & 869 & 1,257\\
    libjpeg & 1,022 & 94 & 651 & 60 & 67 & 2\\
    libxml & 1,596 & 517 & 392 & 16 & n/a$^\dagger$ & 370\\
    mupdf & 487 & 370 & 371 & 38 & n/a & 142\\
    zlib & 376 & 374 & 371 & 15 & 362 & 256\\
    harfbuzz & 6,081 & 3,255 & 4,021 & 111 & n/a & 2,724\\
  \bottomrule
\end{tabular}
{\scriptsize \dag indicates cases where Klee failed to run due to external dependencies}
\end{table}
We collect the edge coverage information from \afl's edge coverage report. 
The results are summarized in Table~\ref{tab:cov}. 
For all $10$ real-world programs, \sys significantly outperforms other fuzzers in terms of edge coverage. 
As shown in Fig~\ref{fig:nn_vs_AFL}, \sys can achieve significantly more new edge coverage than other fuzzers within the first hour. On programs \texttt{strip}, \texttt{harfbuz} and \texttt{readelf}, \sys can achieve more than $1,000$ new edge coverage \textit{within an hour}. For programs \texttt{readelf} and \texttt{objdump}, the number of new edge coverage from \sys's 1 hour running even beats the numbers of new edge coverage from all other fuzzers' 24 hours running. This shows the superior edge coverage ability of \sys.
For all 9 out of 10 programs, \sys achieves 6$\times$,1.5$\times$,9$\times$,1.8$\times$,3.7$\times$,1.9$\times$,10$\times$,1.3$\times$ and 3$\times$ edge coverage than baseline \afl, respectively, and 4.2$\times$,1.3$\times$,7$\times$,1.2$\times$,2.5$\times$,1.5$\times$,1.5$\times$,1.3$\times$ and 3$\times$ edge coverage than the second highest number among all $6$ fuzzers. For the smallest program \texttt{zlib}, which has less than 2k lines of code, \sys achieves similar edge coverage with other fuzzers. We believe it reaches a saturation point when most of the possible edges for such a small program are already discovered after 24 hours fuzzing. The significant outperformance shows the effectiveness of \sys in efficiently locating and mutating critical bytes using the gradient to cover new edges. {\sys} also scales well in large systems. In fact, for programs with more than 10K lines (\eg, 
 \texttt{readelf}, \texttt{harfbuzz}, \texttt{mupdf} and \texttt{libxml}),
 \sys achieves the highest edge coverage, where the taint-assisted fuzzer (\ie, \vuzzer) and symbolic execution assisted fuzzer (\ie, \kleefl) either perform badly or does not scale. 

The gradient-guided mutation strategy allows \sys to explore diverse edges, while other evolutionary-based fuzzers often get stuck and repetitively check the same branch conditions. Also, the minimal execution overhead of the NN smoothing technique helps \sys to scale well for larger programs while other advanced evolutionary fuzzers incur high execution overhead due to the use of heavyweight program analysis techniques like taint-tracking or symbolic execution. 

Among the evolutionary fuzzers, 
\aflfast, uses an optimized seed selection strategies that focuses more on rare edges and thus achieves higher coverage than \afl on $8$ programs, especially in \texttt{libjpeg}, \texttt{size} and \texttt{harfbuzz}. 
\vuzzer, on the other hand, achieves higher coverage than \afl, AFLFast, and \lafintel within the first hour on small programs (\eg, \texttt{zlib}, \texttt{nm}, \texttt{objdump}, \texttt{size} and \texttt{strip}), but its lead stalls quickly and eventually is surpassed by other fuzzers. Meanwhile, \vuzzer's performance degrades on larger programs like \texttt{readelf}, \texttt{harfbuzz}, \texttt{libxml}, and \texttt{mupdf}. We suspect that the imprecisions introduced by \vuzzer's taint tracker causes it to perform poorly on large programs.   
\kleefl uses additional seeds generated by the symbolic execution engine Klee to guide AFL's exploration. Similar to \vuzzer, for small programs (\texttt{nm}, \texttt{objdump}, and \texttt{strip}), \kleefl has good performance at the beginning, but its advantage of additional seeds from Klee fade away after several hours.  Moreover, \kleefl is based on Klee that cannot scale to large programs with complex library code, a well-known limitation of symbolic execution. Thus, \kleefl does not have results on programs \texttt{libxml}, \texttt{mupdf} and \texttt{harfbuzz}. Unlike \vuzzer and \kleefl, \sys does not rely on any heavy program analysis techniques; \sys uses the gradients computed from NNs to generate promising mutations even for larger programs. The efficient NN gradient computation process allow \sys to scale better than \vuzzer and \kleefl at identifying the critical bytes that affect different unseen program branches, achieving significantly more edge coverage. 
 
\lafintel transforms complex magic number comparison into nested byte-comparison using an LLVM pass and then runs \afl on the transformed binaries. It achieves second-highest new edge coverage on program \texttt{strip}. 
However, the comparison transformations add additional instructions to common comparison operations and thus cause a potential edge explosion issue. The edge explosion greatly increases the rate of edge conflict and hurt the performance of evolutionary fuzzing. Also, these additional instructions cause extra execution overheads. As a result, programs like \texttt{libjpeg} with frequent comparison operations suffer significant slowdown (\eg, \texttt{libjpeg}), and \lafintel struggles to trigger new edges.

\RS{2}{\sys can achieve significantly higher edge coverage compared to other gray-box fuzzers (up to 4 $\times$ better than \afl, and 2.5$\times$ better than the second-best one for 24-hour running). 
}
\RQA{3}{\rqc}
\label{req:rnn}
\vspace{-2pt}
Existing recurrent neural network (RNN)-based fuzzers learn mutation patterns from past fuzzing experience to guide future mutations~\cite{rajpal2017not}. 
These models first learn mutation patterns (composed of critical bytes) from a large number of mutated inputs generated by \afl. Next, they use the mutation patterns to build a filter to \afl which only allows mutations on critical bytes to pass, vetoing all other non-critical byte mutations. We choose 4 programs studied by the previous work to evaluate the performance of \sys compared to the RNN-based fuzzer for 1 million mutations. We train two NN models with the same training data, then let the two NN-based fuzzers run to generate 1 million mutations and compare the new code coverage achieved by the two methods. We report both the achieved edge coverage and training time, as shown in~\Cref{tab:cnn_rnn}.

\begin{table}[!htpb]
\setlength{\tabcolsep}{5pt}
\centering
\renewcommand{\arraystretch}{1.1}
  \caption{\textbf{\small \sys vs.~RNN fuzzer \wrt baseline AFL}}
  \label{tab:cnn_rnn}  
  \begin{tabular}{lrrr|rrr}
    \toprule
    \multirow{2}{*}{Programs} & \multicolumn{3}{c|}{\bf Edge Coverage} & \multicolumn{3}{c}{\bf Training Time (sec)} \\\cmidrule{2-7}
     & \sys  & RNN & \afl & \sys & RNN &  \afl\\
    \midrule
    readelf -a &1,800 &215 & 213 & 108 & 2,224 & NA \\
    libjpeg & 89 & 21 & 28 & 56 & 1,028 & NA \\
    libxml & 256 & 38 & 19 & 95 & 2,642 & NA \\
    mupdf & 260 & 70 & 32 & 62 & 848 &  NA \\
  \bottomrule
\end{tabular}
\end{table}

For all the four programs, \sys significantly outperforms the RNN-based fuzzer on 1M mutations. \sys achieves $8.4\times, 4.2\times, 6.7\times$, and $3.7\times$ more edge-coverage than the RNN-based fuzzer across the four programs respectively. 
In addition, the RNN-based fuzzer has, on average, 20$\times$ more training overhead than \sys, because RNN models are significantly more complicated than feed-forward network models. 


An additional comparison of the RNN-based fuzzer with \afl shows that the former achieves $2\times$ more edge coverage on average than \afl on \texttt{libxml} and \texttt{mupdf} using the 1-hour corpus. We also observe that the RNN-based fuzzer vetoes around 50\% of the mutations generated by \afl. Thus, the new edge coverage of 1M mutations from RNN-based fuzzer can achieve the edge coverage of 2M mutations in vanilla \afl. 
This explains why the RNN-based fuzzer uncovers around $2\times$ more new edges of \afl on some programs. If \afl gets stuck after 2M mutations, the RNN-based fuzzer would also get stuck after 1M filtered mutations.
The key advantage of \sys over the RNN-based fuzzer is that \sys obtains critical locations using neural-network-based gradient-guided search, while the RNN fuzzer tries to model the task in an end-to-end manner. Our model can distinguish different contributing factors of critical bytes that the RNN model may miss as demonstrated by our experimental results. For mutation generation, we perform an exhaustive search for critical bytes determined by corresponding contributing factors, while the RNN-based fuzzer still relies on \afl's uniform random mutations.

\RS{3}{\sys, a fuzzer based on simple feed-forward network, significantly outperforms the RNN-based fuzzers by achieving 3.7$\times$ to 8.4$\times$ more edge coverage across different projects.
}

\RQA{5}{\rqd}
\label{req:model}
\sys's fuzzing performance heavily depends on the accuracy of the trained NN. As described in Section~\ref{sec:impl}, we empirically find that an NN model with $1$ hidden layer is expressive enough to model complex branching behavior of real-world programs. In this section, we conduct an ablation study by exploring different model settings for a $1$ hidden layer architecture, \ie, a linear model, an NN model without refinement, and an NN model with incremental refinement. We evaluate the effect of these models on \sys's performance. 

To compare the fuzzing performance, we generate 1M mutations for each version of \sys on $4$ programs. We implement the linear model by removing the non-linear activation functions used in the hidden layer and thus making the whole feed-forward network completely linear. The NN model is trained same seed corpus from \afl. Next, We generate 1M mutations from the passive learning model and measure the edge coverage achieved by these 1M mutations. Finally, we filter out the mutated inputs that exercise unseen edges from the 1 million mutations and add these selected inputs to original seed corpus to incrementally retrain another NN model and use it to generate further mutations. The results are summarized in Table~\ref{tab:model_selection}. We can see that both NN models (with or without incremental learning) outperform the linear models for all $4$ tested programs. This shows that the nonlinear NN models can approximate program behaviors better than a simple linear model. We also observe that incremental learning helps NNs to achieve significantly higher accuracy and therefore higher edge coverage.   

\RS{5}{NN models outperform linear models and incremental learning makes NNs even more accurate over time.
}
\begin{table}[!t]
\setlength{\tabcolsep}{3.5pt}
\centering
\renewcommand{\arraystretch}{1.1}
	\caption{\textbf{\small Edge coverage comparison of 1M mutations generated by \sys using different machine learning models.}}
	\label{tab:model_selection}  
  \begin{tabular}{lrrr}
    \toprule
    Programs & Linear Model &  NN Model & NN + Incremental\\
    \midrule
    readelf -a & 1,723 & 1,800 & 2,020 \\
    libjpeg &  63 & 89 & 159 \\
    libxml & 117 & 256 & 297 \\
    mupdf & 93 & 260 & 329\\
  \bottomrule
\end{tabular}
\end{table}

\section{Case Studies of Bugs}
\label{sec:case_study}

In this section, we provide samples of and analyze three different types of bugs discovered by \sys: integer overflow, out-of-memory, and crash-inducing bugs. 

We note that a large number of program bugs result from incorrect handling of extreme values of variables.
As \sys can enumerate all critical bytes from \texttt{0x00} to \texttt{0xff} (see Algorithm~\ref{alg:mutation} line 3), we manage to find a large number of bugs caused by mishandled extrema. For example, \sys is able to find many out-of-memory bugs in \texttt{libjpeg}, \texttt{objdump}, \texttt{nm} and \texttt{strip} by setting the input bytes that affect memory allocation size to extremely large values. 


%

\noindent\textbf{\texttt{strip}'s integer overflow.}
\sys found an integer overflow bug that can induce an infinite loop on \texttt{strip}. Listing~\ref{lst:strip} shows a function in the \texttt{strip} program that parses every section in the program header table of an input ELF file and assigns all sections to a new program header table in the output ELF file. The integer overflow occurs at the if-condition in line $11$ of Listing~\ref{lst:strip} as \sys sets \texttt{segment\_size} to an extremely large value. Consequently, the program gets stuck in an infinite loop. 
We found that this bug exists in both the latest version of Binutils 2.30 and in older versions 2.26 and 2.29.  



\noindent\textbf{\texttt{libjpeg}'s out-of-memory.}
During the JPEG compression process, the data of every color space is down-sampled by the corresponding sampling factor in order to reduce file size. According to the JPEG standard, the sampling factor must be an integer between 1 and 4. This value is used during the decompression process to determine how much memory needs to be allocated as shown in Listing~\ref{lst:jpeg}. \sys sets a large value which causes too much memory to be allocated for image data, causing a out-of-memory error. Such errors can potentially be exploited to launch denial of service attacks on servers using \texttt{libjpeg} for displaying images.

\lstset{basicstyle=\footnotesize\ttfamily,breaklines=true}
\begin{lstlisting}[caption={\texttt{strip} integer overflow}, captionpos=b, label={lst:strip}, frame=None, belowcaptionskip=.01cm, belowskip=1pt]
// binutils-2.30/bfd/elf.c:6499
#define IS_CONTAINED(saddr, ssize, baddr) \
  (saddr >= baddr
    && saddr <= (baddr + ssize))

rewrite_elf_program_header(bfd *ibfd, bfd *obfd)
{
  for(j = 0; j < section_count; j++)
  {
      output_section = section->output_section;
      if(IS_CONTAINED(output_section,
        segment_size, base_addr))
      {
        ...
        isec++;
        sections[j] = NULL;
        ...
      }
  }
}
\end{lstlisting}

\begin{lstlisting}[caption={\texttt{readelf} section header parsing bug},captionpos=b,label={lst:readelf}, frame=None, belowcaptionskip=.01cm, belowskip=1pt]
// binutils-2.30/binutils/readelf.c:5901
static bfd_boolean 
process_section_headers(Filedata* filedata)
{
  filedata->section_headers = NULL;  
  ...
  if(filedata->file_header.e_shnum == 0) 
  {
    ...
    return TRUE;
  }
}
// binutils-2.30/binutils/readelf.c:654
static Elf_Internal_Shdr *
find_section(Filedata* filedata, char* name)
{
  ...
  assert(filedata->section_headers != NULL);
  ...
}
\end{lstlisting}

\begin{lstlisting}[caption={\texttt{libjpeg} out-of-memory bug}, captionpos=b, label={lst:jpeg}, frame=None, belowskip=1pt]
// libjpeg/jmemmgr.c:444
alloc_barray(j_common_ptr cinfo, ...)
{
  ...
  while (currow < numrows) {
    ...
    alloc_large((size_t) rowsperchunk * (size_t) blocksperrow * SIZEOF(JBLOCK));
    ...
  }   
}
\end{lstlisting}



\noindent\textbf{\texttt{readelf}'s crash.}
An ELF file consists of a file header, program header, section header and section data. According to the ELF specification, the ELF header contains the field \texttt{e\_shnum} located at the 60th byte for a 64-bit binary, which specifies the number of sections in the ELF file. \sys sets the number of sections of the input file to be $0$.   
As shown in Listing~\ref{lst:readelf}, if the number of sections is equal to $0$, the implementation returns a \texttt{NULL} pointer which is dereferenced by subsequent code, triggering a crash.

\section{Related Work}
\label{sec:related}

\noindent\textbf{Program smoothing.} 
Parnas et al.~\cite{parnas1985software} observed that discontinuities are one of the fundamental challenges behind the development of secure and reliable software. Chaudhury et al.~\cite{chaudhuri2011smoothing,chaudhuri2010continuity, chaudhuri2012continuity} suggested the idea of program smoothing to facilitate program analysis and presented a rigorous smoothing algorithm using abstract interpretation and symbolic execution. Unfortunately, such algorithms incur prohibitive performance overhead, especially for large programs. By contrast, our smoothing technique leverages the learning power of NNs to achieve better scalability.

\medskip
\noindent\textbf{Learning-based fuzzing.}
Recently, there has been increasing interest in using machine learning techniques for improving fuzzers~\cite{godefroid2017learn, rajpal2017not, wang2017skyfire, glade, hvlearn, bottinger2018deep, nichols2017faster}. 
However, existing learning-based fuzzers model fuzzing as an end-to-end ML problem, \ie, they learn ML models to directly predict input patterns that can achieve higher code coverage. By contrast, we first use NNs to smoothly approximate the program branching behavior and then leverage gradient-guided input generation technique to achieve higher coverage.  Therefore, our approach is more tolerant to learning errors by ML models than the end-to-end approaches. In this paper, we empirically demonstrate that our strategy outperforms end-to-end modeling both in terms of finding bugs and achieving higher edge coverage~\cite{rajpal2017not}.

\medskip
\noindent\textbf{Taint-based fuzzing.}
Several evolutionary fuzzers have tried to use taint information to identify promising mutating locations~\cite{taintscope, dowser, borg, vuzzer, steelix, angora}. For example, TaintScope~\cite{taintscope} is designed to identify input bytes that affects system/library calls and focus on mutating these bytes. Similarly, Dowser~\cite{dowser} and BORG~\cite{borg} specifically use taint information to target detection of buffer boundary violations and buffer over-read vulnerabilities respectively. By contrast, Vuzzer~\cite{vuzzer} captures magic constants through static analysis and mutates existing values to these constants. Steelix~\cite{steelix} instruments binaries to collect additional taint information about comparing instructions. Finally, Angora~\cite{angora} uses dynamic taint tracking to identify promising mutation locations and perform coordinate descent to guide mutations on these locations. 

However, all these taint-tracking-based approaches are fundamentally limited by the fact that dynamic taint analysis incurs very high overhead while static taint analysis suffers from a high rate of false positives.  Our experimental results demonstrate that \sys easily outperforms existing state-of-the-art taint-based fuzzers by using neural networks to identify promising locations for mutation. 

Several fuzzers and test input generators~\cite{harman2010theoretical, szekeres2017memory, angora} have tried to use different forms of gradient-guided optimization algorithms directly on the target programs. However, without program smoothing, such techniques tend to struggle and get stuck at the discontinuities.

\medskip
\noindent\textbf{Symbolic/concolic execution.}
Symbolic and concolic execution~\cite{king1976symbolic, cadar2008klee, sen2005cute, smart_fuzz, sage} use Satisfiability Modulo Theory (SMT) solvers to solve path constraints and find interesting test inputs. Several projects have also tried to combining fuzzing with such approaches~\cite{symfuzz, kleefl, driller}. Unfortunately, these approaches struggle to scale in practice due to several fundamental limitations of symbolic analysis including path explosion, incomplete environment modeling, large overheads of symbolic memory modeling, etc.~\cite{cadar2013symbolic}. 

Concurrent to our work, NEUEX~\cite{shen2018neuro} made symbolic execution more efficient by learning the dependencies between intermediate variables of a program using NNs and used gradient-guided neural constraint solving together with traditional SMT solvers. By contrast, in this paper, we focus on using NNs to make fuzzing more efficient as it is by far the most popular technique for finding security-critical bugs in large, real-world programs.

\vspace*{.2cm}
\noindent\textbf{Neural programs.} A neural program is essentially a neural network that learns a latent representation of the target program's logic. Several recent works have synthesized such neural programs from input-output samples of a program to accurately predict the program's outputs for new inputs~\cite{graves2014neural, reed2015neural, neelakantan2015neural}. By contrast, we use NNs to learn smooth approximations of a program's branching behaviors.

\section{Conclusion}
\label{sec:conclusion}
We present \sys, an efficient learning-enabled fuzzer that uses a surrogate neural network to smoothly approximate a target program's branch behavior. We further demonstrate how gradient-guided techniques can be used to generate new test inputs that can uncover different bugs in the target program. Our extensive evaluations show that \sys significantly outperforms other $10$ state-of-the-art fuzzers both in the numbers of detected bugs and achieved edge coverage.
Our results demonstrate the vast potential of leveraging different gradient-guided input generation techniques together with neural smoothing to significantly improve the effectiveness of the fuzzing process. 

\section*{acknowledgement}
We thank our shepherd Matthew Hicks and the anonymous reviewers for their constructive and valuable feedback. This work is sponsored in part by NSF grants CNS- 18-42456, CNS-18-01426, CNS-16-17670, CNS-16-18771, CCF-16-19123, CNS-15-63843, and CNS-15-64055; ONR grants N00014-17-1-2010, N00014-16-1- 2263, and N00014-17-1-2788; an ARL Young Investigator (YIP) award; a Google Faculty Fellowship; and a Amazon Web Services grant. Any opinions, findings, conclusions, or recommendations expressed herein are those of the authors, and do not necessarily reflect those of the US Government, ONR, ARL, NSF, Google, or Amazon.

\bibliographystyle{abbrv}
\bibliography{paper}
\balance

\end{document}